\documentclass{article}
\usepackage[top=1in,bottom=1in,left=1in,right=1in]{geometry}
\usepackage{amsmath}
\usepackage{amssymb}
\usepackage{dsfont}
\usepackage{bbold}
\usepackage{times}
\usepackage{array}
\usepackage{makeidx}
\usepackage{graphicx}
\usepackage{subfigure}
\usepackage{bigints}
\usepackage{braket}
\usepackage{tikz}

\usetikzlibrary{backgrounds,fit,decorations.pathreplacing,calc}

\begin{document}

	\title{The Second Law of Quantum Complexity and  \\ the Entanglement Wormhole}
	\author{Andrea Russo\\ \\ Mathematical Tripos - Part III Essay 2018-2019\\
		Department of Applied Mathematics and Theoretical Physics,\\
		Cambridge University.\\ \\
		Under the supervision of: Dr. Johannes Bausch}
	
	\date{}
	\maketitle
\thispagestyle{empty}
	\begin{abstract}
Quantum complexity arises as an alternative measure to the Fubini metric between two quantum states. Given two states and a set of allowed gates, it is defined as the least complex unitary operator capable of transforming one state into the other. Starting with K qubits evolving through a k-local Hamiltonian, it is possible to draw an analogy between the quantum system and an auxiliary classical system with $2^{K}$ degrees of freedom \cite{SecondLaw}. Using the definition of complexity to write a metric for the classical system, it is possible to relate its entropy with the quantum complexity of the K qubits, defining the Second Law of Quantum Complexity. The law states that, if it is not already saturated, the quantum complexity of a system will increase with overwhelming probability towards its maximum value. 
In the context of AdS/CFT duality and the ER=EPR conjecture \cite{ER=EPR}, the growth of the volume of the Einstein Rosen bridge interior, is proportional to the quantum complexity of the instantaneous state of the conformal field theory \cite{Stanford2014} \cite{CAction}. Therefore, the interior of the wormhole connecting two entangled CFT will grow as a natural consequence of the complexification of the boundary state.

	\end{abstract}

\newpage

	\tableofcontents

\newpage

	\section{Introduction}

Quantum complexity was born as a way of measuring a particular kind of distance between two quantum states that the Fubini metric was not able to capture. Chapter II and III will be devoted to define quantum complexity for states and operators and to set a precise dynamics of a quantum system of K qubits. Chapter IV and V will delineate how unitary operators on K qubits can be represented in a classical system with $2^K$ degrees of freedom and how a complexity metric can be imposed on this space. Chapters VI and VII will then exploit the set up of the previous chapters to show how the positional entropy of the classical system can be identified with the quantum complexity of the quantum system, hence explaining the growth of the complexity as a consequence of the growth of entropy. The notion of quantum complexity is then carried in the framework of the AdS/CFT duality. The duality is a vast topic, chapter VIII and IX will introduce the key concepts necessary for the understanding of the connection with quantum complexity. Lastly, chapter X will explain that complexity in AdS/CFT can have two different interpretations. The summary and the discussion are then presented in chapter XI.Throughout this dissertation, unless specified, the convention $c=k_b=\hbar =1$ will be used.
Moreover, whenever the symbol $\approx$ appears, it means that an overall constant has been dropped since the value of interest taken by that quantity is the asymptotical one, reached when one of the variables is really big.

	\section{Quantum Complexity}

		\subsection{Relative quantum complexity}
In this chapter the notion of quantum complexity will be defined. The interesting feature about this quantity is that it will act as a bridge between three very different systems defined in chapter III, IV and XIII. It will be shown that, even if quantum complexity arises purely from quantum computational considerations, its properties allow for a relation between a the evolution of a quantum system and the evolution in a different completely classical non-relativistic system. Moreover, quantum complexity will be found to appear in the domain of high enrgy physics, in particular in the context of the AdS/CFT duality that will be introduced in chapter XIII. Hence, understanding what quantum complexity is and whih are its properties is crucial for appreciating the connection that in recent years seems to be arising between quantum information theory, classical mechanics and high energy physics.
The best way to understand quantum complexity is to start from the question: how far apart are two quantum states? The usual way of measuring this distance is through the Fubini-Study metric. Let $\ket\psi$ and $\ket \phi$ be two normalized quantum states, then

	\begin{equation}
		d_{\phi\psi}=arccos|\braket{\psi|\phi}|.
	\end{equation}

The Fubini metric is therefore bounded $d_{\phi\psi}\in[0,\frac{\pi}{2}]$. When it is very small, the expectation value of all the observables in the states $\ket\psi$ and $\ket \phi$ are close. Consider two different pairs of states. In the first one, the states are a pure tensor product of identical qubits that differ only by one qubit

	\begin{equation}
		\ket\phi=\ket{000...000},\\
		\ket\psi=\ket{000...001}
	\end{equation}

while the second pair is made by pure states in the computational basis which qubits are chosen randomly. 

	\begin{equation}
		\ket\phi'=\ket{01101...1100},\\
		\ket\psi'=\ket{10011...0111}.
	\end{equation}

In both cases the states will be maximally further apart in the Fubini distance

	\begin{equation}
		d_{\phi\psi}=d_{\phi'\psi'}=\frac{\pi}{2}.
	\end{equation}

Even if the two pairs are technically at the same distance, the first pair is closer together in a sense that this inner product fails to capture. This distance has an operational meaning, and it can be understood by thinking about how hard it would be to transition between one state of the pair to the otherthrough the use of unitary gates. Alternatively, one could think about it as a measure of how hard it would be to create a coherent superposition of the states in a pair.  If it was possible to apply any kind of unitary gate, this distinction would be meaningless. However, if only the use of a specific universal set of gates is allowed, then every unitary operation can be decomposed in terms of the allowed set and a new measure of distance can be defined. 

	\begin{description}
\item[\underline{Def}:] Let $\mathcal{U}$ be a universal set of unitary gates and let $\ket\psi$ and $\ket \phi$ be two quantum states. Consider all the possible unitary operators $U$ such that: 
	\begin{equation}
		\ket\psi=U\ket \phi,
	\end{equation}

where for semplicity $U$ is constructed in series as  $U=g_ng_{n-1}...g_1$ with $g_i\in\mathcal{U}$\footnote{Care has to be taken as one is trying to get arbitrarily close to a general unitary operator by composition of smaller gates. The problem can be traced to the Solovay-Kitaev theorem about trying to fill $SU(N)$ with $SU(k)$ gates. Given a tolerance $\epsilon>0$ it is possible to approximate U to a precision $\epsilon$ using $O(log^c (\frac{1}{\epsilon}))$ gates from a fixed finite set, with c depending on the allowed gateset. This can drastically impact the value of the relative complexity. This issue is not investigated deeply in the references because, as it will be clear from the subsequent chapters, the interest is currently mainly on the way complexity grows rather than on its precisel value.}.
The \textit{relative quantum complexity} between $\ket\psi$ and $\ket \phi$ is defined as the circuit complexity $\mathcal{C}(U)$ of the operator U, which is the minimum number of gates that it takes to construct $U$ in this way.

		\begin{equation}
			\mathcal{C}(\ket\psi,\ket\phi):=\mathcal{C}(U)
		\end{equation}
	\end{description}

It is useful to have a symmetric definition of relative complexity, and this will be a necessary condition to define a complexity metric in chapter V. In order to satisfy this requirement, it is enough to allow the Hermitian conjugate of the chosen set of gates to be part of the set.

	\begin{equation}
		\begin{array}{cc}
		 \forall g\in\mathcal{U} &  g^\dag\in\mathcal{U}.
		\end{array}
	\end{equation}

Relative complexity can be also defined between operators. If $V$ and $U$ are two unitary operators, their relative complexity is

	\begin{equation}
		\mathcal{C}(U,V):=\mathcal{C}(UV^\dag)=\mathcal{C}(VU^\dag).
	\end{equation}

It appears that the notion of relative complexity is heavily dependent on the allowed set of gates. For example, only gates acting on 2 or fewer qubits could be permitted. This leads to a family of complexity measures which are all multiplicatively related to each other due to the property that every universal set of gates can be simulated by any other universal set of gates. In particular, every unitary operator on a $K$ qubit system may be written as a product of at most $2^{K-1}(2^K-1)$ two-level unitary matrices.

		\subsection{Properties of Complexity} 

It is useful to list the properties of relative complexity such that they can  be referred to quickly. Let $U,V \in SU(N)$ and let (7) be obeyed, then Complexity satifies:\\

	\begin{enumerate}
\item Non negativity:
		\begin{equation}
			\mathcal{C}(U,V) \geq 0.
		\end{equation}
\item Identity of indiscernibles:
		\begin{equation}
			\mathcal{C}(U,V)=0 \Leftrightarrow U=V
		\end{equation}
\item Symmetry:
		\begin{equation}
			\mathcal{C}(U,V)=\mathcal{C}(V,U).
		\end{equation}
\item The triangle inequality:
		\begin{equation}
			\mathcal{C}(U,V) \leq \mathcal{C}(U,W) + \mathcal{C}(W,V).
		\end{equation}
\item Right invariance:
		\begin{equation}
			\mathcal{C}(U,V)=\mathcal{C}(UR,VR).
		\end{equation}
	\end{enumerate}
It is then possible to see that complexity is a good candidate to be a \textit{right invariant metric}. 

\newpage
	\section{Dynamics of K Qubits}

		\subsection{K-local Hamiltonians}

Now that quantum complexity has been understood, it is possible to proceed in defining the first ine of the triad of systems that will be connected through the notion of quantum complexity. The system is purely quantum and its evolution will follow the rules of quantum mechanics. When the kind of Hamiltonian for this system has to be defined, the choice is taken directly from quantum computational results arisig in black hole physics \cite{BHmirror}. This choice may appear peculiar, but it is crucial in order to be able to later link the notion of quantum complexity with the AdS/CFT duality.
Consider a system of $K$ qubits that share no correlation, this will be referred to as $\mathcal{Q}$. The dynamics of the system is determined by the Hamiltonian, which regulates how qubits evolve and interact. The class of Hamiltonians that will be used are k-local Hamiltonians. Recall that the \textit{weight} $\omega$ of an operator can be defined as the number of single qubit factors that appear in the operator.

	\begin{description}
\item[\underline{Def}:] A \textit{k-local} Hamiltonian is made of building blocks which are Hermitian operators with weight no higher than k. The Hamiltonian is \textit{exactly} k-local if all the operators have $\omega$=k. k-locality does not imply spacial locality, and non-local coupling of qubits is allowed.
	\end{description}
The general form of an exactly k-local Hamiltonian built from standard qubits is:
	\begin{equation}
		H=\sum_{i_1<i_2<...i_k}\sum_{a_1=\{x,y,z\}}...\sum_ {a_k=\{x,y,z\}}J^{a_1,...a_k}_{i_1,...,i_k}\sigma^{a_1}_{i_1}\sigma^{a_2}_{i_2}...\sigma^{a_k}_{i_k}.
	\end{equation}
Or, schematically:
	\begin{equation}
		H=\sum_{I}J_I\sigma_I,
	\end{equation}
where $\sigma_I$ is the set of 3K generalised Pauli operators $\sigma_i^a$, free from spatial locality restrictions so that $I$ runs over all $(4^k-1)$ values, provided that only the k-local couplings are non zero. The properties of the system are considered when averaged over a Gaussian statistical ensemble of J coefficients.
	\begin{equation}
		P(J)=\frac{1}{Z}e^{\frac{1}{2}B_a\sum J_I^2}
	\end{equation}

where the $B_a$ constants define the distribution variance \cite{SecondLaw}.
The reasons behind this choice of operators can be found in \cite{BHmirror}\cite{Jerusalem}. The degrees of freedom of a non-rotating neutral Black Hole can be modelled as thermalized, non-localised qubits living on a stretched horizon hovering at one Planck length $\textit{l}_p$ from the Schwartzchild horizon\footnote{The idea of the streched horizon arises from the need of a theory for which the effective number of degrees of freedom goes to zero very close to the horizon. This is needed in order to fix the near-horizon divergence in the entropy modes of a free QFT caused by the divergence of the local temperature. Under the hypothesis of the exixtence of a free QFT which is adequate down to a distance scale of $\epsilon$, the inequality $\epsilon^2\lesssim G\hbar$ is found. Degrees of freedom are sparse or nonexistent for $\epsilon<\sqrt{G\hbar}$ suggesting the replacement of the mathematical horizon with a an effective membrane called the 'stretched horizon' at a distance of roughly one planck lenght from the mathematical horizon. A physical system in this region is now timelike instead of nulllike, allowing for its dynamical evolution and processes such as the thermalisation of perturbations.}. Every bit of information dropped on the horizon will create a local perturbation, disturbing the thermal equilibrium. Black Holes have no hair, therefore the perturbation will spread out until the thermal equilibrium is reached again. The qubits of the perturbation get \textit{scrambled} with the other degrees of freedom. Exactly 2-local Hamiltonians are a good way of modelling this thermalization process, they are known as 'Fast Scramblers' \cite{Chaos}. Fast scramblers are currently undergoing a lot of studies due to their many interesting properties.  In particular, they can achieve the scrambling time $t_*$, necessary for a perturbation to thermalize, in a logarithmic number of steps. This will be clarified after the circuit model of 2-local Hamiltonians is presented in 3.3. Since scrambling has been mentioned, it needs to be pointed out that, in the literature, different definitions of scrambling are presented. Here the choice reflects the one made in \cite{Chaos}, which uses the notion of \textit{Page scrambling}.

	\begin{description}
\item[\underline{Def}:] Consider a complex chaotic system N with n degrees of freedom prepared in a pure state. The system is \textit{scrambled} if any subsystem smaller than half of the degrees of freedom has maximum entanglement entropy. If m is the number of degrees of freedom of the subsystem $M\subset N$, with $m<\frac{n}{2}$, then
		\begin{equation}
			S(M)=log(m)
		\end{equation}
so that N can be defined as scrambled.
	\end{description}

This means that the information is completely mixed in the degrees of freedom and not available unless at least half of the system is analysed.

		\subsection{The circuit model}
Given a time independent k-local Hamiltonian, the evolution of the quantum system can be implemented through the unitary operator
	\begin{equation}
		U(t)=exp(-iHt).
	\end{equation}

This k-local circuit is believed to be a fast scrambler and the circuit dieagram can be constructed as follows. For a system of K qubits and a 2-local Hamiltonian, at each time step, the qubits are paired randomly and each pair interacts by a randomly chosen gate from a universal set. After each time step, the qubits are regrouped randomly and each new pair interacts again.

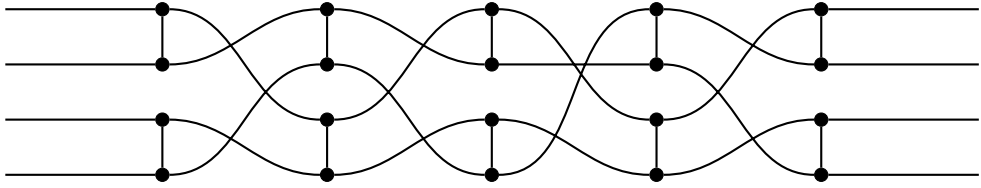
\begin{figure}[h]
	\begin{center}
\begin{tikzpicture}{thick}
\tikzstyle{phase} = [draw,fill,shape=circle,minimum size=5pt,inner sep=0pt]
\matrix[row sep=0.5cm, column sep=2cm] (circuit) {
\node (q1) {}; &
\node[phase] (P11) {}; &
\node[phase] (P12) {}; &
\node[phase] (P13) {}; &
\node[phase] (P14) {}; &
\node[phase] (P15) {}; &

\coordinate (end1); \\
\node (q2) {}; &
\node[phase] (P21) {}; &
\node[phase] (P22) {}; &
\node[phase] (P23) {}; &
\node[phase] (P24) {}; &
\node[phase] (P25) {}; &

\coordinate (end2); \\
\node (q3) {}; &
\node[phase] (P31) {}; &
\node[phase] (P32) {}; &
\node[phase] (P33) {}; &
\node[phase] (P34) {}; &
\node[phase] (P35) {}; &

\coordinate (end3); \\
\node (q4) {}; &
\node[phase] (P41) {}; &
\node[phase] (P42) {}; &
\node[phase] (P43) {}; &
\node[phase] (P44) {}; &
\node[phase] (P45) {}; &

\coordinate (end4); \\
};
\begin{pgfonlayer}{background}
\draw[thick] (q1) -- (P11) to[out= 0, in=180] (P32) to[out= 0, in=180] (P13) to[out= 0, in=180] (P34) to[out= 0, in=180] (P15) -- (end1) 
		(q2) -- (P21) to[out= 0, in=180] (P12) to[out= 0, in=180] (P23) to[out= 0, in=180] (P24) to[out= 0, in=180] (P45)-- (end4)
		(q3) -- (P31) to[out= 0, in=180] (P42) to[out= 0, in=180] (P33) to[out= 0, in=180] (P44) to[out= 0, in=180] (P35)-- (end3)
		(q4) -- (P41) to[out= 0, in=180] (P22) to[out= 0, in=180] (P43) to[out= 0, in=180] (P14) to[out= 0, in=180] (P25)-- (end2)    
		(P11) -- (P21)  (P12) -- (P22) (P13) -- (P23) (P14) -- (P24) (P15) -- (P25)
		(P31) -- (P41)  (P32) -- (P42) (P33) -- (P43) (P34) -- (P44) (P35) -- (P45);
\end{pgfonlayer}
\end{tikzpicture}
\end{center}
\caption{2-local Hamiltonian acting on 4 qubits, at each step the qubits are paired and interact through a random generalised Pauli matrix}
\end{figure}

At each timestep, $\frac{K}{2}$ gates act in parallell. Hence, to prepare the unitary $U(\tau)$ the number of gates acting is 
	\begin{equation}
		N_{gates}(\tau)=\frac{K}{2}\tau.
	\end{equation}
$\tau$ is referred as the \textit{depth} of the circuit and K as the \textit{width}. Since it is believed \cite{BHmirror}\cite{NegCurv} that at least for some lenght of time this circuit is the most efficient, the complexity grows linearly as:
	\begin{equation}
		\mathcal{C}[U(\tau)]=\frac{K}{2}\tau.
	\end{equation}
		
		\subsection{Scrambling and Switchback}

Let's see how the scrambling time is achieved. Consider a perturbation given by a single-qubit operator $W$, which could represent a photon with energy $\delta E=\frac{1}{8\pi MG}$ falling into the Black Hole. The choice of energy is such that the entropy of the Black Hole increases by one unit. To see its effect on the system, it is useful to introduce the concept of a precursor \cite{NegCurv}.

	\begin{description}
\item[\underline{Def}:] Let W be a single qubit operator representing a perturbation of the system. A \textit{precursor} is an operator of the form 
		\begin{equation}
			W(\tau)=U(\tau)WU^\dag(\tau).
		\end{equation}
Where $U(\tau)$ is the time evolution operator (11).
	\end{description}

$W(\tau)$ applied at a time $t=0$ represents the action of a perturbation applied at $t=-\tau$ spreading through the system. This is achieved because the operator $U(\tau)^\dag$ runs time backwards before the perturbation is applied and then $U(\tau)$ moves time forward again to the moment at which one is interested in measuring the effect of the perturbation.
The effect of a precursor is better understood by looking at the circuit model.

\begin{tikzpicture}{thick}
\tikzstyle{phase} = [draw,shape=circle,minimum size=5pt,inner sep=0pt]
\matrix[row sep=0.2cm, column sep=0.45cm] (circuit) {
\node (q1) {}; &
\node[phase,fill] (P11) {}; &
\coordinate (P12) {}; &
\node[phase,fill] (P13) {}; &
\coordinate (P14) {}; &
\coordinate(P15) {}; &
\coordinate(P16) {}; &
\coordinate (P17) {}; &
\node[phase,fill] (P18) {}; &
\coordinate (P19) {}; &
\coordinate (P110) {}; &
\node[phase] (P111) {}; &
\coordinate (P112) {}; &
\coordinate (P113) {}; &

\coordinate (P114) {}; &
\node[phase] (P115) {}; &
\coordinate (P116) {}; &
\coordinate (P117) {}; & %
\node[phase,fill] (P118) {}; &
\coordinate (P119) {}; &
\coordinate(P120) {}; & %
\coordinate(P121) {}; &
\coordinate (P122) {}; &
\node[phase,fill] (P123) {}; &
\coordinate (P124) {}; &
\node[phase,fill] (P125) {}; &
\coordinate (end1);\\

\node (q2) {}; &
\node[phase,fill] (P21) {}; &
\coordinate (P22) {}; &
\coordinate (P23) {}; &
\node[phase,fill] (P24) {}; &
\coordinate (P25) {}; & 
\coordinate(P26) {}; &
\node[phase,fill] (P27) {}; &
\coordinate (P28) {}; &
\coordinate (P29) {}; &
\node[phase] (P210) {}; &
\coordinate (P211) {}; &
\coordinate (P212) {}; &
\coordinate (P213) {}; &

\coordinate (P214) {}; &
\coordinate (P215) {}; &
\node[phase] (P216) {}; &
\coordinate (P217) {}; &
\coordinate (P218) {}; &
\node[phase,fill] (P219) {}; &
\coordinate(P220) {}; &
\coordinate (P221) {}; & 
\node[phase,fill] (P222) {}; &
\coordinate (P223) {}; &
\coordinate (P224) {}; &
\node[phase,fill] (P225) {}; &
\coordinate (end2);\\

\node (q3) {}; &
\node[phase,fill] (P31) {}; &
\coordinate (P32) {}; &
\coordinate (P33) {}; &
\node[phase,fill] (P34) {}; &
\coordinate (P35) {}; &
\coordinate(P36) {}; &
\node[phase,fill] (P37) {}; &
\coordinate (P38) {}; &
\coordinate (P39) {}; &
\coordinate (P310) {}; &
\coordinate (P311) {}; &
\node[phase,fill] (P312) {}; &
\coordinate (P313) {}; &

\node[phase,fill] (P314) {}; &
\coordinate (P315) {}; &
\coordinate (P316) {}; &
\coordinate (P317) {}; &
\coordinate (P318) {}; &
\node[phase,fill] (P319) {}; &
\coordinate(P320) {}; &
\coordinate (P321) {}; &
\node[phase,fill] (P322) {}; &
\coordinate (P323) {}; &
\coordinate (P324) {}; &
\node[phase,fill] (P325) {}; &
\coordinate (end3);\\

\node (q4) {}; &
\node[phase,fill] (P41) {}; &
\coordinate (P42) {}; &
\coordinate (P43) {}; &
\coordinate (P44) {}; &
\node[phase,fill] (P45) {}; &
\coordinate(P46) {}; &
\node[phase] (P47) {}; &
\coordinate (P48) {}; &
\coordinate (P49) {}; &
\node[phase] (P410) {}; &
\coordinate (P411) {}; &
\coordinate (P412) {}; &
\coordinate (P413) {}; &

\coordinate (P414) {}; &
\coordinate (P415) {}; &
\node[phase] (P416) {}; &
\coordinate (P417) {}; &
\coordinate (P418) {}; &
\node[phase] (P419) {}; &
\coordinate(P420) {}; &
\node[phase,fill] (P421) {}; &
\coordinate (P422) {}; &
\coordinate (P423) {}; &
\coordinate (P424) {}; &
\node[phase,fill] (P425) {}; &
\coordinate (end4);\\

\node (q5) {}; &
\node[phase,fill] (P51) {}; &
\coordinate (P52) {}; &
\node[phase,fill] (P53) {}; &
\coordinate (P54) {}; &
\coordinate (P55) {}; &
\coordinate(P56) {}; &
\node[phase] (P57) {}; &
\coordinate (P58) {}; &
\coordinate (P59) {}; &
\coordinate (P510) {}; &
\node[phase] (P511) {}; &
\coordinate (P512) {}; &
\coordinate (P513) {}; &

\coordinate (P514) {}; &
\node[phase] (P515) {}; &
\coordinate (P516) {}; &
\coordinate (P517) {}; &
\coordinate (P518) {}; &
\node[phase] (P519) {}; &
\coordinate(P520) {}; &
\coordinate (P521) {}; &
\coordinate (P522) {}; &
\node[phase,fill] (P523) {}; &
\coordinate (P524) {}; &
\node[phase,fill] (P525) {}; &
\coordinate (end5);\\

\node (q6) {}; &
\node[phase,fill] (P61) {}; &
\coordinate (P62) {}; &
\coordinate (P63) {}; &
\coordinate (P64) {}; &
\node[phase,fill] (P65) {}; &
\coordinate(P66) {}; &
\coordinate (P67) {}; &
\node[phase,fill] (P68) {}; &
\coordinate (P69) {}; &
\coordinate (P610) {}; &
\coordinate (P611) {}; &
\node[phase,fill] (P612) {}; &
\node[draw,rectangle,fill] (P613) {}; &

\node[phase,fill] (P614) {}; &
\coordinate (P615) {}; &
\coordinate (P616) {}; &
\coordinate (P617) {}; &
\node[phase,fill] (P618) {}; &
\coordinate (P619) {}; &
\coordinate(P620) {}; &
\node[phase,fill] (P621) {}; &
\coordinate (P622) {}; &
\coordinate (P623) {}; &
\coordinate (P624) {}; &
\node[phase,fill] (P625) {}; &
\coordinate (end6);\\
};
	\begin{pgfonlayer}{background}
\draw            (q1)  -- (end1) (q2)--(end2) (q3)--(end3) (q4)--(end4) (q5)--(end5) (q6)--(end6)
		(P12) -- (P62) (P16)--(P66) (P19)--(P69) (P113)--(P613) (P117)--(P617) (P120)--(P620) (P124)--(P624)
		(P11)--(P21) (P31)--(P41) (P51)--(P61)
		(P13)--(P53) (P24)--(P34) (P45)--(P65)
		(P18)--(P68) (P27)--(P37) (P47)--(P57)
		(P111)--(P511) (P210)--(P410) (P312)--(P612)
		(P115)--(P515) (P314)--(P614) (P216)--(P416)
		(P118)--(P618) (P219)--(P319) (P419)--(P519)
		(P123)--(P523) (P421)--(P621) (P222)--(P322)
		(P125)--(P225) (P325)--(P425) (P525)--(P625)
		(P613) node[below=0.1cm]{W};
	\end{pgfonlayer}
\end{tikzpicture}

The effect of the inserted operator spreads through all the other qubits as the perturbation thermalizes with the rest of the system, but operators acting on unaffected qubits cancel out with their mirror gate.

\begin{tikzpicture}{thick}
\tikzstyle{phase} = [draw,shape=circle,minimum size=5pt,inner sep=0pt]
\matrix[row sep=0.2cm, column sep=0.48cm] (circuit) {
\node (q1) {}; &
\node[phase,fill] (P11) {}; &
\coordinate (P12) {}; &
\node[phase,fill] (P13) {}; &
\coordinate (P14) {}; &
\coordinate(P15) {}; &
\coordinate(P16) {}; &
\coordinate (P17) {}; &
\node[phase,fill] (P18) {}; &
\coordinate (P19) {}; &
\coordinate (P110) {}; &
\coordinate (P111) {}; &
\coordinate (P112) {}; &
\coordinate (P113) {}; &

\coordinate (P114) {}; &
\coordinate (P115) {}; &
\coordinate (P116) {}; &
\coordinate (P117) {}; & %
\node[phase,fill] (P118) {}; &
\coordinate (P119) {}; &
\coordinate(P120) {}; & %
\coordinate(P121) {}; &
\coordinate (P122) {}; &
\node[phase,fill] (P123) {}; &
\coordinate (P124) {}; &
\node[phase,fill] (P125) {}; &
\coordinate (end1);\\

\node (q2) {}; &
\node[phase,fill] (P21) {}; &
\coordinate (P22) {}; &
\coordinate (P23) {}; &
\node[phase,fill] (P24) {}; &
\coordinate (P25) {}; & 
\coordinate(P26) {}; &
\node[phase,fill] (P27) {}; &
\coordinate (P28) {}; &
\coordinate (P29) {}; &
\coordinate (P210) {}; &
\coordinate (P211) {}; &
\coordinate (P212) {}; &
\coordinate (P213) {}; &

\coordinate (P214) {}; &
\coordinate (P215) {}; &
\coordinate (P216) {}; &
\coordinate (P217) {}; &
\coordinate (P218) {}; &
\node[phase,fill] (P219) {}; &
\coordinate(P220) {}; &
\coordinate (P221) {}; & 
\node[phase,fill] (P222) {}; &
\coordinate (P223) {}; &
\coordinate (P224) {}; &
\node[phase,fill] (P225) {}; &
\coordinate (end2);\\

\node (q3) {}; &
\node[phase,fill] (P31) {}; &
\coordinate (P32) {}; &
\coordinate (P33) {}; &
\node[phase,fill] (P34) {}; &
\coordinate (P35) {}; &
\coordinate(P36) {}; &
\node[phase,fill] (P37) {}; &
\coordinate (P38) {}; &
\coordinate (P39) {}; &
\coordinate (P310) {}; &
\coordinate (P311) {}; &
\node[phase,fill] (P312) {}; &
\coordinate (P313) {}; &

\node[phase,fill] (P314) {}; &
\coordinate (P315) {}; &
\coordinate (P316) {}; &
\coordinate (P317) {}; &
\coordinate (P318) {}; &
\node[phase,fill] (P319) {}; &
\coordinate(P320) {}; &
\coordinate (P321) {}; &
\node[phase,fill] (P322) {}; &
\coordinate (P323) {}; &
\coordinate (P324) {}; &
\node[phase,fill] (P325) {}; &
\coordinate (end3);\\

\node (q4) {}; &
\node[phase,fill] (P41) {}; &
\coordinate (P42) {}; &
\coordinate (P43) {}; &
\coordinate (P44) {}; &
\node[phase,fill] (P45) {}; &
\coordinate(P46) {}; &
\coordinate (P47) {}; &
\coordinate (P48) {}; &
\coordinate (P49) {}; &
\coordinate (P410) {}; &
\coordinate (P411) {}; &
\coordinate (P412) {}; &
\coordinate (P413) {}; &

\coordinate (P414) {}; &
\coordinate (P415) {}; &
\coordinate (P416) {}; &
\coordinate (P417) {}; &
\coordinate (P418) {}; &
\coordinate (P419) {}; &
\coordinate(P420) {}; &
\node[phase,fill] (P421) {}; &
\coordinate (P422) {}; &
\coordinate (P423) {}; &
\coordinate (P424) {}; &
\node[phase,fill] (P425) {}; &
\coordinate (end4);\\

\node (q5) {}; &
\node[phase,fill] (P51) {}; &
\coordinate (P52) {}; &
\node[phase,fill] (P53) {}; &
\coordinate (P54) {}; &
\coordinate (P55) {}; &
\coordinate (P56) {}; &
\coordinate (P57) {}; &
\coordinate (P58) {}; &
\coordinate (P59) {}; &
\coordinate (P510) {}; &
\coordinate (P511) {}; &
\coordinate (P512) {}; &
\coordinate (P513) {}; &

\coordinate (P514) {}; &
\coordinate (P515) {}; &
\coordinate (P516) {}; &
\coordinate (P517) {}; &
\coordinate (P518) {}; &
\coordinate (P519) {}; &
\coordinate(P520) {}; &
\coordinate (P521) {}; &
\coordinate (P522) {}; &
\node[phase,fill] (P523) {}; &
\coordinate (P524) {}; &
\node[phase,fill] (P525) {}; &
\coordinate (end5);\\

\node (q6) {}; &
\node[phase,fill] (P61) {}; &
\coordinate (P62) {}; &
\coordinate (P63) {}; &
\coordinate (P64) {}; &
\node[phase,fill] (P65) {}; &
\coordinate(P66) {}; &
\coordinate (P67) {}; &
\node[phase,fill] (P68) {}; &
\coordinate (P69) {}; &
\coordinate (P610) {}; &
\coordinate (P611) {}; &
\node[phase,fill] (P612) {}; &
\node[draw,rectangle,fill] (P613) {}; &

\node[phase,fill] (P614) {}; &
\coordinate (P615) {}; &
\coordinate (P616) {}; &
\coordinate (P617) {}; &
\node[phase,fill] (P618) {}; &
\coordinate (P619) {}; &
\coordinate(P620) {}; &
\node[phase,fill] (P621) {}; &
\coordinate (P622) {}; &
\coordinate (P623) {}; &
\coordinate (P624) {}; &
\node[phase,fill] (P625) {}; &
\coordinate (end6);\\
};
	\begin{pgfonlayer}{background}
\draw            (q1)  -- (end1) (q2)--(end2) (q3)--(end3) (q4)--(end4) (q5)--(end5) (q6)--(end6)
		(P12) -- (P62) (P16)--(P66) (P19)--(P69) (P113)--(P613) (P117)--(P617) (P120)--(P620) (P124)--(P624)
		(P11)--(P21) (P31)--(P41) (P51)--(P61)
		(P13)--(P53) (P24)--(P34) (P45)--(P65)
		(P18)--(P68) (P27)--(P37) 
		(P312)--(P612)
		(P314)--(P614)
		(P118)--(P618) (P219)--(P319)
		(P123)--(P523) (P421)--(P621) (P222)--(P322)
		(P125)--(P225) (P325)--(P425) (P525)--(P625)
		(P613) node[below=0.1cm]{W};

	\end{pgfonlayer}
\end{tikzpicture}

The result is a complexity which is less than the total number of gates obtained by simply counting the gates in $UWU^\dag$
	\begin{equation}
		\mathcal{C}(UWU^\dag)\leq \mathcal{C}(U) + \mathcal{C}(W) + \mathcal{C}(U^\dag).
	\end{equation}
This is called the \textit{switchback effect}.
If the portion of the K of qubits affected is denoted by $s(\tau)$, then at each step the average number of qubits affected is
	\begin{equation}
		\Delta s=\frac{(K-s)s}{K-1}.
	\end{equation}
Ignoring the 1 at the denominator as K is usually very large and writing this as a differential equation
	\begin{equation}
		\frac{ds}{d\tau}=\frac{(K-s)s}{K}
	\end{equation}
which, when integrated gives the ratio of affected qubits
	\begin{equation}
		\frac{s(\tau)}{K}=\frac{e^{\tau-\tau_*}}{1+e^{\tau-\tau_*}}.
	\end{equation}
As the ratio approches 1, the perturbation gets thermalised. Complete scrambling is reached at the \textit{srambling time} $\tau=\tau_*$.
	\begin{equation}
		\tau_*=log(K).
	\end{equation}
Therefore, a k-local Hamiltonian is a fast scrambler as the number of steps needed is logarithmic in the number of qubits of the system.
To computemore precisely the complexity of the precursor and the magnitude of the switchback effect, consider that the size of the affected qubits is the rate of growth of the precursor complexity
	\begin{equation}
		\frac{d\mathcal{C}(\tau)}{d\tau}=s(\tau)= K\frac{e^{\tau-\tau_*}}{1+e^{\tau-\tau_*}}.
	\end{equation}
Hence, the complexity grows as
	\begin{equation}
		\mathcal{C}(\tau)= Klog(1+e^{\tau-\tau_*}) = \begin{cases} e^\tau & \mbox{if } \tau \ll \tau_* \\ \frac{K}{2}(2\tau-2\tau_*) & \mbox{if } \tau \gg \tau_* \end{cases}
	\end{equation}

	\begin{figure}[h]
		\begin{center}
			\includegraphics[keepaspectratio,scale=0.6]{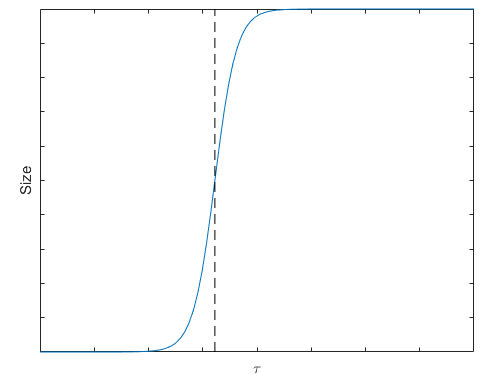}
		\end{center}
			\caption{Growth of the perturbation size. As the time $\tau$ increases on the x axis, the number of qubits affected by the perturbation stays low until the scrambling time $\tau_*$ is reached, when the size of the perturbation grows immediately to its saturation value. }
	\end{figure}
\newpage

	\begin{figure}[h]
		\begin{center}
			\includegraphics[keepaspectratio,scale=0.6]{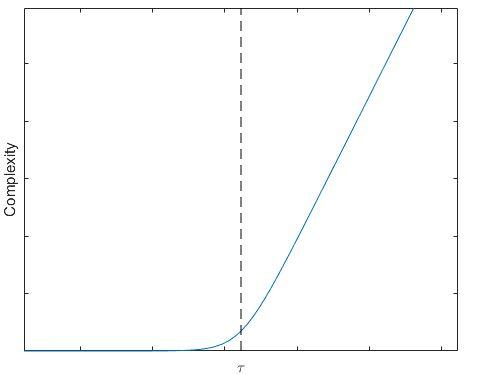}
		\end{center}
			\caption{Switchback effect of the precursor. As the time $\tau$ increases on the x axis, the cmplexity of the precursor grows exponentially until the scrambling time $\tau_*$ is reached, afterwards it grows linearly.}
	\end{figure}

In both figures 2 and 3, the quantities are plotted against the time $\tau$, and the scrambling time $\tau_*$ is indicated by a dashed line.
As it is clear from figure 2, the size of the precursor jumps almost immediately to its maximum at the scrambling time. In figure 3, the complexity grows exponentially until scrambling has occurred and then proceeds to grow linearly as indicated in (27).

\newpage


	\section{A classical model for quantum bits}

		\subsection{The classical auxiliary space}
The second system of the three mentioned before will now be defined. This will be a completely classical environment acting as the background for a non-relativistic classical motion. It will then be linked to the quantum system $\mathcal{Q}$ through the imposition of a complexity-derived  metric.
Unitary operators acting on a system of $K$ qubits are elements of the Lie group $SU(2^K)$ and can be represented as $2^K \times 2^K$ matrices. The group has exactly $4^k-1$ degrees of freedom and its generators are the generalised Pauli matrices $\sigma_I$. The time evolution of an operator $U$ under the influence of a Hamiltonian $H$ is given by

	\begin{equation}
		i\frac{dU}{dt}=HU
	\end{equation}

It is possible to eliminate H by rearranging the terms and derive an equation of motion that is not directly dependent on the Hamiltonian

	\begin{equation}
		\frac{d^2U}{dt^2}-\dot{U}U^\dag\dot{U}=0
	\end{equation}

where the dot represents a time derivative.\\
This recalls the form of the geodesic equation of a non-relativistic particle moving on $SU(N)$

	\begin{equation}
		\frac{d^2x^\rho}{dt^2}+\Gamma^{\rho}_{\mu\nu}\dot{x^\mu}\dot{x^\nu}=0,
	\end{equation}

where $x$ are generalised coordinates and the Christoffel symbols are determined by the metric imposed on the space.
Let's call this classical system $\mathcal{A}$. The aim is to define a metric on $\mathcal{A}$ derived from the definition of relative complexity between operators, such that the motion of the fictitious particle given by (30) will trace paths which length will be connected to the relative complexity between the operators at the starting and final point. The origin of $\mathcal{A}$ is taken to be the identity operator.

	\subsection{Maximum complexity}

Prior to defining the complexity metric for the space $\mathcal{A}$, it is useful to understand what is the maximum complexity for unitary operators acting on K qubits. As a starting point let's consider an approximation of the number of unitary operators in SU(N). According to \cite{Volume} the volume of SU(N) is 

	\begin{equation}
		Vol(SU(N))\approx \prod_{k=1}^{N-1}\frac{2\pi^{k+1}}{k!}.
	\end{equation}
Since 
	\begin{equation}
		\sum_{k=1}^{N-1}(k+1)=\sum_{k=1}^{N-1}k + (N-1)= \frac{(N-1)(N+2)}{2}
	\end{equation}
then
	\begin{equation}
		Vol(SU(N))\approx\frac{2\pi^{\frac{(N+2)(N-1)}{2}}}{1!2!....(N-1)!}.
	\end{equation}

Recalling that any unitary operator can be approximated with an error $\epsilon$ by a universal finite set of gates, and that
SU(N) has dimension $N^2-1$, the volume of an $\epsilon$-ball is:

	\begin{equation}
		V_n(\epsilon)=\frac{\pi^\frac{n}{2}}{\Gamma(\frac{n}{2}+1)}\epsilon^n,
	\end{equation}
using
	\begin{equation}
		\Gamma(n)=(n-1)!
	\end{equation}
the volume of a $N^2-1$ dimensional ball becomes:
	\begin{equation}
		V_{N^2-1}(\epsilon)=\frac{\pi^\frac{N^2-1}{2}}{(\frac{N^2-1}{2})!}\epsilon^{N^2-1}.
	\end{equation}
The number of unitaries in SU(N) can be roughly estimated by identifying every operator with an $\epsilon$-ball. Keeping in mind that for N large: $N^2-1\approx N^2$, and using Stirling formula
for approximating factorials:
	\begin{equation}
		ln(n!)\approx nln(n)-n
	\end{equation}
it is possible to have an estimate of the number of unitaries dividing (34) by (37)
	\begin{equation}
		\mathcal{N}\approx \left( \frac{N}{\epsilon^2}\right) ^\frac{N^2}{2} \Rightarrow  \left( \frac{2^K}{\epsilon^2}\right) ^\frac{4^K}{2}.
	\end{equation}
It is insightful to look at the logarithm of this quantity
	\begin{equation}
		ln(\mathcal{N})\approx \frac{4^K}{2}Kln(2) + 4^K ln\left( \frac{1}{\epsilon}\right),
	\end{equation}

which exhibits a strong dependence on the number of qubits K but a weak dependence on the chosen error tolerance $\epsilon$.
Let's now analyse the evolution of K qubits by starting with the identity operator and acting upon it with a sequence of circuits of depth one. Without loss of generality, the allowed set of gates $\mathcal{U}$ will be composed by a single non-symmetric 2-gate. At each application, the qubits must be paired and acted upon. The number of different choices in the first step is

	\begin{equation}
		d\approx\frac{K!}{(K/2)!}\approx \left( \frac{2K}{e} \right) ^\frac{K}{2}.
	\end{equation} 
This represents \textit{d} different unitary operators that can be thought as points in $SU(2^K)$ which are one 'complexity step' further from the identity operator.\\
It is now necessary to make two assumptions. The first is that, at each subsequent step, the choice of pairings is not the same as in the previous step. This amounts to say that in further steps the number of choices is

	\begin{equation}
		\frac{K!}{(K/2)!}-1.
	\end{equation}

The second and most important assumption is that no collisions occur. This means that for a fixed level of depth in the circuit, the probability of two unitary operators to be the same is vanishingly small. The assumption is equivalent to say that it is very rare for two different paths in $\mathcal{A}$ to get within the same $\epsilon$-ball. This assumption will eventually break down at late times as the complexity approaches its maximum, but it is safe at early times and can be justified by the exponentially high number of $\epsilon$-balls contained in $SU(2^k)$.\\
With these assumptions, the number of unitaries reached at depth D is

	\begin{equation}
		\mathcal{N}_D=d^D\approx \left( \frac{2K}{e} \right)^{\frac{KD}{2}}.
	\end{equation}
Since no collisions occur, the path to each unitary is minimal and hence the depth times the width of the circuit represents the complexity of the operator.
	\begin{equation}
		\mathcal{C}(U) = \frac{K}{2}D.
	\end{equation}
Which agrees with the linear growth of (20) after the identification $D=\tau$.
Therefore, the number of operators with a given complexity grows exponentially as the complexity increases
	\begin{equation}
		\mathcal{N}(\mathcal{C}) \approx \left( \frac{2K}{e} \right)^{\mathcal{C}}.
	\end{equation}
Eventually, maximum complexity is reached, and this happens when all the operators of $SU(2^K)$ can be reached by following a path form the identity. The circuit has then complexity $\mathcal{C}_{max}$
	\begin{equation}
		\left( \frac{2K}{e} \right)^{\mathcal{C}_{max}} = \left( \frac{2^K}{\epsilon^2}\right) ^{\frac{4^K}{2}},
	\end{equation}
which gives
	\begin{equation}
		\mathcal{C}_{max} = 4^K \left[ \frac{1}{2} + \frac{|log(\epsilon)|}{log(K)} \right].
	\end{equation}
Again it is possible to notice the strong dependence of the complexity on the number of qubits K and its weak dependence on the dimension of the $\epsilon$-ball,
	\begin{equation}
		\mathcal{C}_{max}\approx 4^K.
	\end{equation}

\newpage
	\section{The Complexity Metric}

		\subsection{Continuous complexity metric}

It is now time to use the notions developed in the previous chapters to impose a complexity metric on the space $\mathcal{A}$. In chapter II it was observed that relative complexity satisfies the properties of a right invariant metric. Right invariant metrics on Riemann manifolds are parametrised by a symmetric 'moment of inertia' tensor \cite{Rightinv} so that the metric takes the form:

	\begin{equation}
		ds^2=\mathcal{I}_{ij}d\Omega_Id\Omega_J.
	\end{equation}
where
	\begin{equation}
		d\Omega_I=iTr(dU^\dag\sigma_IU).
	\end{equation}

In the circuit model, the complexity proceeds by discrete steps governed by the Hamiltonian which coefficients $J_I$ forbid undesired couplings. In the system $\mathcal{A}$ the motion is continuous and the constraints have to be imposed through the tensor $\mathcal{I}_{IJ}$. The idea is to penalise motion along non k-local directions in such a way that the choice of allowed gates $\mathcal{U}$ is respected. The ambiguity in the choice of $\mathcal{I}_{IJ}$ corresponds to the freedom in the choice of $\mathcal{U}$. Moreover, motion in this new 'complexity space' needs to replicate fundamental features like the \textit{switchback effect} and the \textit{scrambling time}. It was shown in \cite{NegCurv} that in order to achieve these behaviours the space needs to have negative sectional curvature of order $1/K$. Intuitively, this can be understood by thinking about how neighbouring geodesics diverge in negatively curved space. The spreading of a perturbation measured by a precursor is represented by initially close paths quickly diverging from each other. Therefore, care has to be taken when choosing the penalties for non k-local directions. In \cite{Nielsen} the penalty was taken to always be the maximum possible complexity $4^K$, but this led to a sectional curvature of order $4^k$, which is incompatible with the other required features for complexity space.\\ 
The metric is hene defined as:

	\begin{equation}
		\mathcal{I}_{IJ} = \delta_{IJ}\mathcal{I}(\omega_I),
	\end{equation}
where no sum is implied.The penalties are set by $\mathcal{I}(\omega)$, where $\omega$ is the weight of the generalised Pauli operator $\sigma_I$
	\begin{equation}
		\mathcal{I}(\omega) = \begin{cases} 1 & \mbox{if } \omega \leq k \\ c4^{\omega-k} & \mbox{if } \omega > k \end{cases}.
	\end{equation}
The penalty factor is independent of the number of qubits and relies on the assumption that the price paid for moving in a specific direction is dependent only on the weight of the Pauli matrices.

		\subsection{Negative curvature}

To see how this metric imposes a negative curvature on the auxiliary system, consider a 2-dimensional section generated by all the geodesics which arises from the action of two 2-local Hamiltonians passing through the origin. Without loss of generality the two Hamiltonians will be $H$ and $H+\Delta d\theta$,

	\begin{equation}
		\begin{array}{ccc}		
 			H & = & \sum J^{\alpha\beta}_{ij}\sigma^\alpha_i \sigma^\beta_j \\
			\Delta & = & \sum D^{\alpha\beta}_{ij}\sigma^\alpha_i \sigma^\beta_j
		\end{array}
	\end{equation}

 where $\Delta$ is a 2-local operator orthogonal to $H$ such that

	\begin{equation}
		Tr(\Delta H)=Tr(H\Delta)=0.
	\end{equation}

Since the surface will not generally have zero extrinsic curvature, strictly speaking, the curvature of this section should be defined at the origin t=0. To generally extend the curvature as a complete function of time, it is necessary to work in leading order of $d\theta$, otherwise geodesics will take shortcuts off the surface.

	\begin{figure}[h]
		\begin{center}
			\includegraphics[keepaspectratio,scale=0.8]{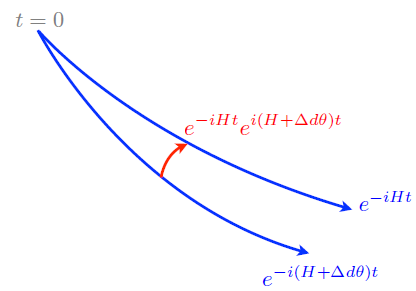}
		\end{center}
			\caption{2-local Hamiltonians generating diverging geodesics. Credits: \cite{SecondLaw}}
	\end{figure}

Two geodesics will evolve from the the origin according to $e^{-iHt}$ and $e^{-i(H+\Delta d\theta)t}$. As the two geodesics evolve, they are connected by a Jacobi field, and the Jacobi vector grows with the time evolution as the paths diverge. The acceleration of the growth indicates the geodesic deviation.

Define the Loschmidt operator as a mean to move between the geodesics

	\begin{equation}
		e^\Lambda=e^{-iHt}e^{-i(H+\Delta d\theta)t}.
	\end{equation}

Recalling the Baker-Cambell-Hausdorff formula for $e^Z=e^Xe^Y$ with $[X,Y]\neq 0$

	\begin{equation}
		\Lambda = -\sum_{m=0}\frac{(it)^{m+1}}{(m+1)!} [\underbrace{H,[H,[H,[H}_\text{m},\Delta]]]]d\theta.
	\end{equation}

Keeping only terms of order $t^3$ and $d\theta$ gives

 	\begin{equation}
		\Lambda= -\left( i\Delta t - \frac{t^2}{2}[H,\Delta] - \frac{it^3}{6}[H,[H,\Delta]] \right) d\theta.
	\end{equation}

Along the direction defined by $\Delta$, the distance is measured by

	\begin{equation}
		dl^2_\bot=Tr[\Lambda^\dag \cdot \Lambda]d\theta^2,
	\end{equation}

while the distance along the radial direction is 

	\begin{equation}
		dl^2_\parallel=Tr[H^\dag \cdot H]dt^2.
	\end{equation}

The dot represents the fact that the product has to be weighted by the appropriate factor of $\mathcal{I}(\omega)$, depending on the k-locality of the operators involved.
Carrying out the calculation explicitly 

	\begin{equation}
		\Lambda^\dag= -\left(- i\Delta t + \frac{t^2}{2}[H,\Delta] + \frac{it^3}{6}[H,[H,\Delta]] \right) d\theta.
	\end{equation}

which can be easily obtained by remembering that the commutator $[H,\Delta]$ is anti-hermitian $[H,\Delta]^\dag=-[H,\Delta]$.
Keeping only terms up to and including $t^4$

	\begin{equation}
		dl^2_\bot=Tr \left[ (\Delta \cdot \Delta)t + \frac{it^3}{2}\Delta \cdot [H,\Delta] - \frac{t^4}{6}\Delta \cdot [H,[H,\Delta]] + \frac{it^3}{2}\Delta \cdot [H,\Delta]
				- \frac{t^4}{4}[H,\Delta] \cdot [H,\Delta] - \frac{t^4}{6}\Delta \cdot [H,[H,\Delta]] \right] d\theta^2.
	\end{equation}

When the trace is computed, the terms of order $t^3$ disappear, regardless of the weight of the operators

	\begin{equation}
		Tr(\Delta\cdot[H,\Delta])=\alpha Tr(\Delta[H,\Delta])=\alpha Tr[\Delta (H\Delta-\Delta H)] = \alpha (Tr[\Delta H\Delta] - Tr[\Delta\Delta H]) = 0.
	\end{equation}

Hence, the total distance is

	\begin{equation}
	ds^2=dl^2_\parallel+dl^2_\bot=Tr[H\cdot H]dt^2 + \left( Tr[\Delta\cdot\Delta]t^2 - \frac{t^4}{3}Tr[\Delta\cdot[H,[H,\Delta]]] - \frac{t^4}{4}Tr[[H,\Delta]\cdot[H,\Delta]]\right)d\theta^2.
	\end{equation}

In order to evaluate the weighted traces, it is necessary to understand the k-locality of the different products. 2-local operators are not weighted by the metric as the motion along their direction is encouraged, operators of higher weight get instead a factor of $\mathcal{I}(\omega)$. Therefore, $Tr[H\cdot H]=Tr[H^2]$ and $Tr[\Delta\cdot\Delta]=Tr[\Delta^2]$ as they are 2-local. The second trace argument gives 2-local and 4-local terms

	\begin{equation}
		Tr(\Delta\cdot[H,[H,\Delta]]) = 2( Tr[\Delta^2H^2] - Tr[\Delta H\Delta H])
	\end{equation}

but the 4-local terms are eliminated when traced against $\Delta$.
The last term contains 1-local and 3-local operators. However, as argued in \cite{SecondLaw}, for a large number $K$ of qubits only the 3-local terms dominate, so it is reasonable to weight the product by $\mathcal{I}(3)$. Therefore,

	\begin{equation}
ds^2=dl^2_\parallel+dl^2_\bot=Tr[H^2]dt^2 + \left( Tr[\Delta^2]t^2 - \frac{t^4}{3}Tr[\Delta[H,[H,\Delta]]] - \frac{t^4}{4}\mathcal{I}(3)Tr[[H,\Delta][H,\Delta]]\right)d\theta^2.
	\end{equation}

Using the identity $ Tr[[H,\Delta][\Delta,H]] = Tr[\Delta[H,[H,\Delta]]]$, the sectional curvature at t=0 is proportional to the coeffient of the $t^4$ with the opposite sign.

	\begin{equation}
		\mathcal{ R}|_{t=0,K\gg k=2}= \left( \frac{1}{3}-\frac{\mathcal{I}(3)}{4}\right) \frac{2Tr([H,\Delta][\Delta,H])}{Tr[\Delta^2]Tr[H^2]}
	\end{equation}

where the numerator is positive due to the anti-hermitian properties of the commutators.
For large $K$, after averaging over the Gaussian coefficients, only a fraction of $1/K$ terms in $H$ does not commute with $\Delta$, giving

	\begin{equation}
		 \frac{2Tr([H,\Delta][\Delta,H])}{Tr[\Delta^2]Tr[H^2]} \sim \frac{1}{K}.
	\end{equation}

Therefore,

	\begin{equation}
		\mathcal{ R}|_{t=0,K\gg k=2} = \frac{2^8}{K} \left( \frac{1}{3}-\frac{\mathcal{I}(3)}{4}\right) + \mathcal{O}\left( \frac{1}{K^2} \right),
	\end{equation}

which is negative for $\mathcal{I}(3)>4/3$.\\
By looking at this formula it is possible to see that the space is negatively curved with the right curvature if the penalty towards non k-local directions is imposed as in (52). Moreover, if the penalty is not gradual but immediately exponential as in \cite{Nielsen}, the sectional curvature will be exponential too, which causes a violent divergence of neighbouring geodesic, nullifying the switchback effect (28) and not reproducing the correct scrambling time (26).

\newpage
	\section{Motion in Complexity space}

		\subsection{Initial conditions and conserved quantities}

Now that the classical auxiliary system $\mathcal{A}$ has been endowed with a complexity metric capable of reproducing the notion of relative complexity and its properties, it is time to study the motion of the fictitious classical particle representing a unitary operator evolving in time. Recalling that the path in $SU(N)$ is traced by the geodesic equation (30) it is possible to see that there is no direct reference to the Hamiltonian. The information is instead encoded in the initial conditions. By substituting the explicit expression for the Hamiltonian (15) into the Schroedinger equation (29)

	\begin{equation}
		\dot{U}=-i\sum_IJ_I\sigma_IU
	\end{equation}

which we can solve for the coefficients $J_I$

	\begin{equation}
		J_I=iTr[\sigma_I\dot{U}U^\dag].
	\end{equation}

At this point, the generality of equation (30) is lost and the time evolution of operators in SU(N) has beenmade to represent the time evolution operator for the $K$ qubits of the system $\mathcal{Q}$.
Since the aim is to investigate the initial conditions, it is possible to set $U=\mathbb{1}$ at the origin to get

	\begin{equation}
		J_I=iTr[\sigma_I\dot U],
	\end{equation}

which means that the coefficients $J_I$ are the projection of the initial velocity onto the axes oriented along the Pauli basis in the tangent space, therefore representing the initial velocity components of the fictitious particle.

	\begin{equation}
		J_I=V_I|_{t=0}.
	\end{equation}

A free non-relativistic particle in geodesic motion has a conserved kinetic energy 

	\begin{equation}
		E_a=\frac{1}{2}V_a^2
	\end{equation}

where the subscript $a$ indicates that the quantities refer to the auxiliary system $\mathcal{A}$.
The variance of the Hamiltonian, using the normalised trace such that $Tr(\mathbb{1})=1$ is

	\begin{equation}
		(\Delta H)^2=Tr(H^2)=Tr\sum_I\sum_J\sigma_I\sigma_J J_I J_J.
	\end{equation}
Since 
	\begin{equation}
		Tr[\sigma_I\sigma_J]=\delta_{IJ},
	\end{equation}
then
	\begin{equation}
		(\Delta H)^2 = \sum_I J_I^2.
	\end{equation}

Since all the terms in the Hamiltonian are traceless, the average of the Hamiltonian is zero. However, the energy E relative to the ground state is not, and it carries the same variance of H.
In chapter III it was explained that the choice of a 2-local Hamiltonian was made in order to be able to draw a connection with static neutral black holes. It is a general property of this kind of black holes that their dimensionless Rindler Energy and its variance are both proportional to the Black Hole entropy, which in turn is also the number of qubits needed to model its internal dynamics. Therefore,

	\begin{equation}
		(\Delta E)^2= \sum_I J_I^2 |_{avg} = E \approx S_{Bek} = K.
	\end{equation}

where the average refers to the ensemble average.

	 	\subsection{Action in complexity space}

The connection between the complexity of the time evolution operator U acting on the quantum system $\mathcal{Q}$ and the geodesic motion of a non-relativistic particle in the auxiliary system $\mathcal{A}$ takes the form of the complexity-action correspondence given by

	\begin{equation}
		\mathcal{C}= \frac{1}{2}\int_{\mathbb{1}}^U \mathcal{I}_{IJ} \dot{X}^I \dot{X}^J  d\tau,
	\end{equation}

where $\mathcal{I}_{IJ}$ is the complexity metric. The equation is then constrained so that the conserved energy $E_a$ is proportional to the actual energy of the system $\mathcal{Q}$

	\begin{equation}
		E_a=K=E.
	\end{equation}

Therefore \cite{SecondLaw},

	\begin{description}
\item[\underline{Def}:] The complexity of a unitary operator U is the minimum action of any trajectory connecting U and the identity, subject to the condition that the energy of the fictitious particle in motion in the auxiliary system is fixed and equal to the number of qubits K.
	\end{description}

The relation between the complexity and the length of the path is obtained using the relation between the action $\mathcal{S}$ and the path length

	\begin{equation}
		\mathcal{S}=\sqrt{E_a}Lenght=\sqrt{K}\Delta L
	\end{equation}

such that

	\begin{equation}
		\Delta\mathcal{C}=\sqrt{K}\Delta L.
	\end{equation}

It is possible now to study the growth of complexity by looking at the ordinary non-relativistic connection between the kinetic energy of a free particle and the Lagrangian of the system $\mathcal{A}$, which in this case gives

	\begin{equation}
		L_a=\frac{1}{2}V_a^2=K.
	\end{equation}

Hence, the rate of growth of the complexity is the number of qubits, and by substituting in (78) the complexity becomes

	\begin{equation}
		\mathcal{C}=\frac{K}{2}\tau
	\end{equation}

which is in agreement with the circuit model complexity (20).

		\subsection{Ergodicity of motion}

Before moving to the formulation of the Second Law of Quantum Complexity, it is important to talk about the ergodicity of the free particle motion in SU(N).
The question is if the motion generated by the exactly 2-local time-independent Hamiltonian (15) will fill the space $SU(2^K)$. The answer is no. To visualise this write the Hamiltonian time evolution unitary operator in the energy basis

	\begin{equation}
		U(\tau)=e^{-iH\tau}= \sum_{i=1}^{2^k}e^{-iE_i\tau}\ket{E_i}\bra{E_i}.
	\end{equation}

Therefore, for a given Hamiltonian the evolution is restricted to a torus defined by the $2^k$ phases

	\begin{equation}
		e^{i\theta_i}=e^{iE_i\tau}.
	\end{equation}

The motion in the $4^K-1$ dimensional system $\mathcal{A}$ is bounded inside a $2^K$ dimensional embedded torus.
As the evolution on the torus is chaotic, the motion is ergodic. The time for U to return in a neighbourhood of the identity is the time needed for all the phases to simultaneously get close to 1, this implies a doubly exponential recurrence time in the number of qubits

	\begin{equation}
		\tau_{recur}\approx e^{2^K}.
	\end{equation}

Does this change if a variation is taken over all possible k-local Hamiltonians of the form (14)? The answer is negative since the Hamiltonians are specified by the parameters J. The number of J parameters for an exactly k-local Hamiltonian is

	\begin{equation}
		N_J=3^k {{K}\choose{k}} \approx \frac{(3K)^k}{k!}
	\end{equation}

which is polynomial in K. Therefore, the dimension of the set covered by a k-local Hamiltonian is only slightly bigger than a $2^K$ dimensional subset.
At this point is already possible to delineate how the complexity-entropy correspondence will play out. As shown in (45), the number of accessible operators for a given complexity grows exponentially. If the number of accessible operators is treated as the number of microstates, the corresponding entropy will be

	\begin{equation}
		S \approx \mathcal{C}log(K).
	\end{equation}

\newpage
	\section{The II law of Quantum Complexity}

		\subsection{Entropy and Complexity}

The molecular chaos hypothesis states that for a classical non-relativistic gas it is possible to separate the phase space probability into a position and momentum component

    \begin{equation}
        P(x,p)=F(x)G(p).
    \end{equation}

As a consequence of this factorisation, the entropy also factorises and can be expressed as the sum of two terms, namely the \textit{positional entropy} and the \textit{kinetic entropy} associated with the position and momentum of the gas

	\begin{equation}
		S= - \int F(x)logF(x)dx - \int G(p)logG(p)dp.
	\end{equation}

Since motion in the $\mathcal{A}$ system is related to the complexity of operators acting on the K qubits of $\mathcal{Q}$ through the complexity metric, the conjecture proposed in \cite{SecondLaw} aims to define a clear relation between the two systems.

    \begin{description}
\item[\underline{Conjecture}:] At any instant, the ensemble average of the computational complexity of the quantum system $\mathcal{Q}$, is proportional to the classical positional entropy of the auxiliary system $\mathcal{A}$.
    \end{description}

Two things need to be highlighted about this conjecture. The first is that the correspondence is between positional entropy and complexity and not total entropy and complexity. This happens because it is only the position of an operator U with respect to the identity $\mathbb{1}$ that indicates its complexity. Moreover, there is a proportionality relation between the two quantities. This is caused by the fact that the complexity depends on a number of factors such as the allowed gate set. It is assumed that differences in the complexity value will show only as a multiplicative ambiguity. The conjecture is supported by the counting argument leading to (88). The number of accessible unitaries grows exponentially, the identification of this quantity with the number of microstates with complexity $\mathcal{C}$ supports the claim that

	\begin{equation}
		\mathcal{C}_\mathcal{Q} \propto S_{\mathcal{A}(position)}.
	\end{equation}

It is important to stress that the positional entropy of the auxiliary system $\mathcal{A}$ is not a physical quantity and it is not related with the entropy of the qubits of the system $\mathcal{Q}$.

		\subsection{Complexity equilibrium and recurrence time}

By looking at how entropy and complexity evolve, it is possible to notice that they both grow linearly until their maximum value is reached. Entropy saturates after a polynomial time in the degrees of freedom which means that complexity needs an exponential time in the number of qubits. The two quantities will then fluctuate around the equilibrium. Fluctuations around maximal complexity can be intuitively understood by thinking that the application of random 2-gates may lead to a small decrease in the complexity due to some cancellations between gates, but the enormous number of operators accessible for a nearly maximal complexity state is so big that the decrease in complexity will not be significant until the recurrence time is reached.
Since the motion in the embedded torus submanifold is ergodic and bounded, Poincar\' e recurrence theorem assures that, eventually, the state of the system will return close to the initial value. The recurrence time for entropy is exponential in the degrees of freedom. Since the degrees of freedom of the system $\mathcal{A}$ are exponential in the number of qubits, the recurrence time needed for complexity to spontaneously decrease to a value close to zero is doubly exponential in the number of qubits.

	\begin{equation}
		t_{\mathcal{C},recur} \approx exp(2^K).
	\end{equation}

\begin{figure}[h]
		\begin{center}
			\includegraphics[keepaspectratio,scale=0.8]{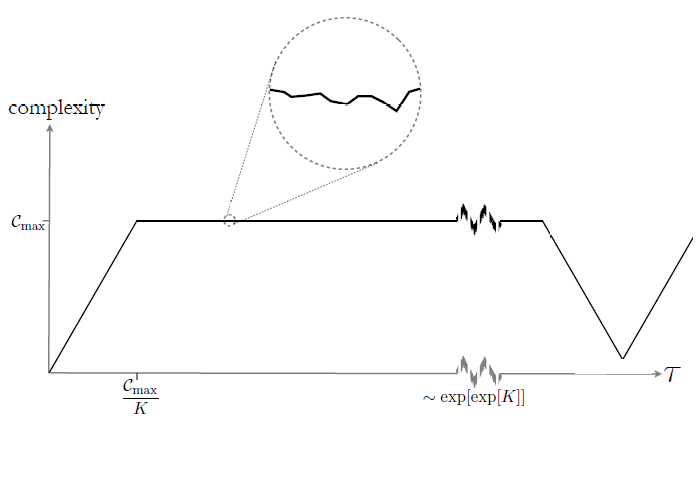}
		\end{center}
			\caption{Complexity evolution and recurrence time. Credits: \cite{SecondLaw}}
	\end{figure}

		\subsection{The Second Law of Quantum Complexity}

The complexity of the quantum system $\mathcal{Q}$ and the entropy of the auxiliary classical system $\mathcal{A}$ are related. Entropy will obey the second law of thermodynamics and its growth towards saturationis understood as a statistically overwhelming probable event. In the same way, it is possible to formulate an equivalent second law for quantum complexity.

	\begin{description}
\item[\underline{Def}:] If the quantum complexity is less than maximum, then with overwhelming likelihood it will increase, both into the future and into the past.
	\end{description}

The growth of entropy and complexity are statistical phenomena. In principle, the time evolution of the system can be inverted if the sign of the Hamiltonian is changed. In classical mechanics, this leads to a decrease in entropy due to the inversion of the phase space trajectories. However, this apparent violation of the second law of thermodynamics is unstable if the system is chaotic as in this case. A small perturbation affecting just a single degree of freedom will spread quickly to the rest of the system and reverse the decrease of entropy. If the same line of reasoning is applied to the auxiliary entropy of $\mathcal{A}$ it is possible to see that even if a system is prepared so that its complexity reduces with time, the complexity decrease is unstable. A perturbation in the form of a single qubit operator will affect the rest of the system after a scrambling time (26), reversing the process and igniting the growth of complexity.

		\subsection{Kolgromov complexity}

According to (90), entropy for the auxiliary system $\mathcal{A}$ factorises in a positional and a kinetic part. The positional entropy is related to the quantum complexity and one would expect the kinetic part to also have a corresponding relation with complexity. Indeed, kinetic entropy is related to Kolgromov complexity \cite{SecondLaw}. Since the kinetic energy is set by the J coefficients of the Hamiltonian and the kinetic entropy of $\mathcal{A}$ can be identified with the entropy of the probability distribution P(J), this complexity is a property of the Hamiltonian and not of the quantum state. For simplicity, imagine that all the J are either 0 or 1. The Hamiltonian (15) could then be identified by a string (011001...). Kolgromov complexity measures the length of the shortest algorithm that can prepare a string, hence the Hamiltonian will have an associated fixed Kolgromov complexity. In the real case, the J coefficients are real numbers, which means that to specify them with infinite accuracy will take an infinite amount of information. Fortunately, this is not a problem. Kolgromov complexity is time independent for a time-independent Hamiltonian like (15). Its value is of the order of the number of J coefficients, scaling linearly with the number of qubits K. This means that it acts as a fixed complexity overhead having to do with the complexity of the algorithm and becomes irrelevant after a short amount of time. At complexity equilibrium, positional entropy, and therefore computational complexity, vastly dominate over the kinetic entropy and Kolgromov complexity.

\newpage
	\section{AdS/CFT in a nutshell}

		\subsection{Dualities}

In physics,  a duality appears when there are two equivalent theories capable of describing the same physical system. The two theories are then interchangeable and can be put in relation through the use of the system they describe. This allows for a better understanding of the theories and of the underlying physics. In the last century, the study of black hole thermodynamics gave rise to the Information Paradox, which resolution prompted the birth of the Holographic principle and consequently the formulation of a nonperturbative description for a quantum gravity theory known as AdS/CFT duality \cite{Informationparadox}. The duality is a vast topic and discussing the details of its formulation is outside of the scope of this work. In this chapter, it will be briefly presented through the explanation of its components, namely Anti de Sitter spacetimes and Conformal Field Theories, and the relation between the two will be delineated focusing on the aspects that are necessary to draw the connection with quantum complexity.

		\subsection{Anti de Sitter spacetimes}

In the absence of a stress-energy tensor, n-dimensional General Relativity is described by the variation of the Einstein-Hilbert action

	\begin{equation}
		\mathcal{S}_{EH} = \int \frac{ (\mathcal{R} - 2\Lambda)}{16\pi G} \sqrt{g}d^nx
	\end{equation} 

giving Einstein equation

	\begin{equation}
		\mathcal{R}_{\mu\nu} - \frac{1}{2}\mathcal{R}g_{\mu\nu} + \Lambda g_{\mu\nu}=0,
	\end{equation}

where $\Lambda$ is the \textit{cosmological constant}.The solutions to this equation are called \textit{vacuum solutions}. The properties of the associated spacetimes depend on the value of the cosmological constant. There are three types of  maximally symmetric solutions depending on the sign of $\Lambda$

	\begin{equation}
		 \begin{array}{clc} 
		\Lambda >0 & \mbox{de Sitter spacetime} & \mbox{positive curvature}\\ 
		\Lambda =0 & \mbox{Minkowsky spacetime} & \mbox{no curvature} \\ 
		\Lambda <0 & \mbox{Anti de Sitter spacetime} & \mbox{negative curvature}
		\end{array}.
	\end{equation}

N-dimensional Anti de Sitter spacetimes are negatively curved and can be embedded in a (n+2)-dimensional flat spacetime with two time dimensions. The embedding is done through the hyperboloid constraint

	\begin{equation}
		-W^2 - U^2 +\sum_{i=1}^{n} X_i^2 = -l_{ads}^2,
	\end{equation}

where $l_{ads}$ is the \textit{radius of cuvature} defined as

	\begin{equation}
		\Lambda = \frac{-(n-1)(n-2)}{2l_{ads}^2}.
	\end{equation}

The manifold can be parametrised with global coordinates $(\tau,\rho,\theta,\phi_1....\phi_{n-3})$ through the mapping

	\begin{equation}
		\begin{array}{ccc} 
			W & = & l_{ads} \cosh(\rho) \cos(\tau) \\ 
			U & = & l_{ads} \cosh(\rho) \sin(\tau) \\
			X_i & = & sinh(\rho) x_i 
		\end{array}
	\end{equation}

where $\sum_i x_i^2=1$ as the $x_i$ parametrise the (n-2)-sphere through 

	\begin{equation}
		\begin{array}{ccc}
			x_1 & = & \sin(\theta)\sin(\phi_1)...\sin(\phi_{n-3}) \\
			x_2 & = & \sin(\theta)\sin(\phi_1)...\cos(\phi_{n-3}) \\
			x_3 & = & \sin(\theta)\sin(\phi_1)...\cos(\phi_{n-2}) \\
			etc.
		\end{array}
	\end{equation}

The metric takes the form

	\begin{equation}
		ds^2=l_{ads}^2(-cosh(\rho)^2d\tau^2 + d\rho^2 +sinh(\rho)^2 d\Omega^2_{n-2}),
	\end{equation}

where $d\Omega_{n-2}$ is the line element of the (n-2)-sphere. Here, $\tau\in[0,2\pi]$ and $\rho\in \mathds{R}^+$. In order to avoid closed timelike curves, the universal cover  $\tau\in\mathds{R}$ is taken.
All n-dimensional Anti de Sitter spacetimes have the topology of $\mathds{R}\times S^{n-1}$. Their symmetry group can be seen from (96) to be $SO(2,n-1)$ and they can be expressed as the quotient of the special orthogonal groups

	\begin{equation}
		AdS_n= \frac{SO(2,n-1)}{SO(1,n-1)}.
	\end{equation}

The boundary of Anti de Sitter Spacetimes is of particular interest for discussing the AdS/CFT correspondence. Let's take 5-dimensional AdS spacetime, for reasons that will be clear later. Its symmetry group is $SO(2,4)$ and its metric can be written as 

	\begin{equation}
		ds^2= \frac{l_{ads}^2}{(1-r^2)^2} \left( -(1+r^2)^2 dt^2 + 4dr^2 + 4r^2 d\Omega^2_3 \right),
	\end{equation}

where $r\in [0,1)$. The centre of the spacetime is at  $r=0$ while infinity is represented by the singularity at $r=1$. As $r\rightarrow 1$ the metric can be approximated by 

	\begin{equation}
		ds^2\approx \frac{4l_{ads}^2}{(1-r^2)^2} \left( -dt^2 + dr^2 + d\Omega^2_3 \right),
	\end{equation}

meaning that the boundary is conformally flat. Moreover, for outgoing radial null geodesics 

	\begin{equation}
		(1+r^2)^2dt^2 = 4dr^2.
	\end{equation}

Therefore, a light ray can reach the conformal boundary and come back in a finite coordinate time. All these properties are summarised in the statement that AdS has a timelike conformal boundary that makes the spacetime into a reflecting cavity. This comes particularly handy when a black hole is inserted in the picture. 
The asymptotically Anti de Sitter metric for a Schwartzchild black hole can be written as

	\begin{equation}
		ds^2 = -\left( 1 - \frac{\mu}{r^{d-3}} + \frac{r^2}{l_{ads}^2} \right) dt^2 + \frac{dr^2}{\left( 1 - \frac{\mu}{r} + \frac{r^2}{l_{ads}^2} \right)} + r^2 d\Omega^2_{d-2}.
	\end{equation}

where $d$ is the number of bulk spacetime dimensions and 

	\begin{equation}
		\mu = \frac{16\pi GM}{(d-2)\Omega_{d-2}}
	\end{equation}

with $\Omega_{d-2}$ the volume of a unit (d-2)-dimensional sphere. 
Usually, when dealing with a black hole in asymptotically flat spacetimes, the assumption is made that the black hole is stable and eternal. This does not hold once Hawking radiation enters the picture. The black hole will gradually lose its mass and eventually disappear. In order to stabilise it, one could think of immersing the black hole in a thermal bath, like a gas, to compensate for the loss of energy. This canonical ensemble cannot work as the temperature of the black hole is inversely proportional to its mass, giving the black hole a negative heat capacity.

	\begin{equation}
		T_{Hawking} = \frac{1}{8\pi MG} 
	\end{equation}

If the black hole temperature is less than the bath, it will absorb some energy, hence lowering its temperature even further. If the black hole is hotter, it will radiate away its mass, raising its temperature. A black hole in equilibrium with the bath is a completely unstable system. A small fluctuation from equilibrium gives rise to a runaway process that either consumes the black hole completely or makes it grow indefinitely. On the other hand, having a black hole in thermodynamic equilibrium is possible if the surrounding spacetime is Anti de Sitter. The reflecting boundary allows the Hawking radiation to act as a finite heat reservoir by being reabsorbed in the black hole. The stability of the AdS-Schwartzchild solution depends on the radius of the black hole and of the kind of ensemble that is being considered. In the context of AdS/CFT the ensemble considered is the canonical ensemble and the black hole is stable if its Schwartchild radius is comparable to the AdS radius of curvature $r_{Schwartz} \sim l_{ads}$ \cite{Hawking1983}. This kind of black holes are stable configurations. The Penrose diagram for a Schwartzchild-AdS spacetime is presented here:

	\begin{figure}[h]
		\begin{center}
			\includegraphics[keepaspectratio,scale=0.55]{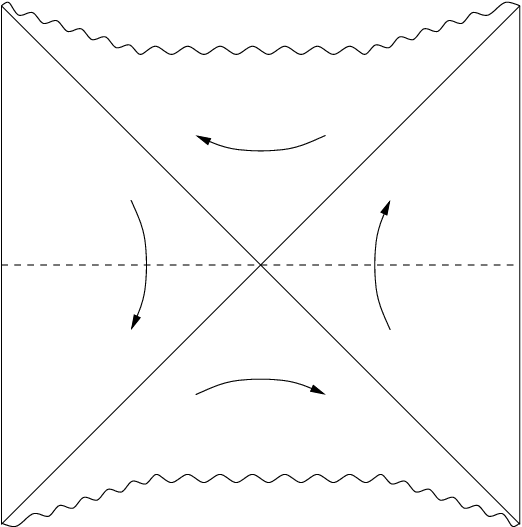}
		\end{center}
			\caption{Penrose diagram of Schwartzchild black hole in AdS. The arrows indicate the coordinate Killing time direction.}
	\end{figure}
\newpage

		\subsection{Conformal Field Theories}

A conformal field theory is a relativistic quantum field theory which is symmetric under a group of spacetime symmetries called the conformal group. The conformal group is an extension of the Poincar\'e group that includes a larger set of angle-preserving transformations. The conformal group has the following representations

	\begin{equation}
		\begin{array}{ccl}
			M_{\mu\nu} & \equiv & i(x_\mu \partial_\nu - x_\nu\partial_\mu) \\
			P_\mu & \equiv & -i\partial_\mu \\
			D & \equiv & -ix_\mu\partial^{\mu} \\
			K_\mu & \equiv & i(x^2\partial_\mu - 2x_\mu x_\nu \partial^{\nu}) 
		\end{array}
	\end{equation}

where $M_{\mu\nu}$ are the Lorentz generators, $P_\mu$ generates translations, $D$ changes the scale of spacetime through dilatations and $K_\mu$ generates the conformal transformations given by

	\begin{equation}
		x'^{\mu}= \frac{x^\mu - a^\mu x^2}{1 - 2a\cdot x + a^2x^2},
	\end{equation}

where $a^\mu$ is a constant vector parametrising the transformation.
In order to preserve the conformal symmetry, CFT cannot have a preferred length scale. This implies that there can't be anything in the theory like a mass or a Compton wavelength. Therefore, when dealing with conformal field theories, usually the interest is not on S-matrices and particle scattering, but rather on correlations functions of operators and their behaviour under conformal transformations. The algebra of the conformal group is isomorphic to the algebra of SO(2,d), hence it is already possible to see similarities between this kind of theories and AdS spacetimes. In any CFT there is a particular set of operators whose commutators with the special conformal generators are zero at the origin \cite{Jerusalem}. These are called \textit{primary operators} and they transform simply under conformal transformations. Under dilatations of the form

	\begin{equation}
		x'^\mu=\lambda x^\mu
	\end{equation}

with $\lambda\in\mathds{R}$, they transform as

	\begin{equation}
		\mathcal{O}'(x')= \lambda^{-\Delta}\mathcal{O}(x),
	\end{equation}

where $\Delta$ is the \textit{conformal scaling dimension} of $\mathcal{O}$. For a free field theory, $\Delta$ can be determined by dimensional analysis of the Lagrangian, for scalar operators it obeys $\Delta \geq \frac{d-2}{2}$ where $d$ is the number of spacetime dimensions. The scaling dimension of a composite operator obtained by taking the product of two operators of dimensions $\Delta_1$ and $\Delta_2$ is $\Delta_1 + \Delta_2$. Primary operators have simple correlation functions, the time ordered two-point function for a scalar primary operator $\mathcal{O}$ in any CFT is

	\begin{equation}
		\bra{\Omega}T\mathcal{O}(x,t)\mathcal{O}(0,0)\ket{\Omega} = \frac{1}{( |x|^2 - t^2 + i\epsilon)^\Delta}.
	\end{equation}
Conformal field theories involved in the AdS/CFT correspondence mainly lie on a fixed flat background, in particular they are often studied on the cilinder
	\begin{equation}
		ds^2= -dt^2 + d\Omega^2_{d-1}.
	\end{equation}

		\subsection{AdS/CFT}

The biggest consequence of the Holographic principle \cite{Informationparadox} was that gravity seems to be, as suggested by the name of the conjecture, holographic. This means that given a region of spacetime, the number of degrees of freedom, and hence the maximum information and entropy carried by that region, are proportional to the area of the region's boundary instead of its volume. This is why canonical ways of quantisation fail to work with gravity. Quantising gravity as a local field theory implies that local fluctuations of spacetime act as degrees of freedom in the same way in which local fluctuations of the Electromagnetic field represent its degrees of freedom. The effect is that a massive overcounting occurs and the theory is not renormalisable \cite{LectonGrav}. In 1997, Maldacena proposed a non-perturbative model of quantum gravity in the form of the AdS/CFT correspondence stating the complete equivalence between superstring theory in the bulk of $AdS_5 \times S^5$ and $\mathcal{N}$=4, 3+1 dimensional, SU(N), Super Yang-Mills theory on the boundary of AdS space\cite{Maldacena}. Its modern formulation states \cite{Jerusalem}:

	\begin{description}
\item[\underline{Def}:] Any relativistic conformal field theory on $\mathds{R} \times S^{d-1}$ with metric (113) can be interpreted as a theory of quantum gravity in an asymptotically $AdS \times M$ spacetime where $M$ is some compact manifold that may or may not be trivial.
	\end{description}

Given the two theories, a \textit{dictionary} is built to map quantities in the boundary CFT to quantities in the bulk AdS spacetime and vice-versa. 
The Hilbert space of physical states of the bulk is by definition identical to the CFT Hilbert space. The symmetry generators of the SO(2,d) group of AdS are identified with the conformal symmetry generators of the CFT. This implies that the Hamiltonian is the same for both the physical theories. As a consequence, all the quantities that depend on the Hamiltonian and/or the space of states are identical. For example the thermal partition function or the free energy at finite temperature. As it was argued at the beginning of this section, bulk fields should not exist in a theory of quantum gravity as they carry an excess of degrees of freedom. Nevertheless, in Minkowski space, the states at past and future infinity can be described in terms of free fields. In the same way, in the AdS/CFT correspondence, a local bulk field can make sense if studied at its boundary limit. If $\mathcal{O}$ is a scalar primary CFT operator, then the dictionary states the existence of a bulk scalar field such that

    \begin{equation}
        \lim_{r\to\infty} r^\Delta\phi (t,r,\Omega) \equiv \mathcal{O}(t,\Omega),
    \end{equation}

which means that a bulk field can be extrapolated to a boundary primary operator of conformal dimension $\Delta$ by stripping off a normalisation factor.
When dealing with non-scalar quantities, the unique stress-energy tensor of the CFT is a spin two primary operator of dimension $d$ and its bulk dual is the metric tensor of AdS, which carries the theory of gravity within the boundary of the spacetime.
The conjecture has not been proven and there is not yet a set of necessary and sufficient conditions that can define whether any particular CFT is holographic. Nevertheless, it is believed that having a gravity dual obeying Einstein equations requires CFT with a large number of degrees of freedom and strong coupling. There exist other conditions for the spectrum of states and operators of the CFT which can be roughly expressed as the requirement for the CFT to have only as many low-energy states as gravity would have in an asymptotically AdS spacetime \cite{LectonGrav}.
The AdS/CFT duality has been used to study very different phenomena, from QCD to condensed matter systems. The connection with quantum complexity emerges when the duality is used to study black holes, and in particular their interior. The vacuum state of a CFT corresponds to pure AdS spacetime, while the geometries corresponding to low energy excitations of the boundary are gravitational waves. For high energy states, the corresponding spacetimes can have significantly different topologies and geometries. The thermal states of a boundary CFT are dual to a black hole in the bulk. In particular, if the CFT is defined on a sphere $S^d$, there is a deconfinement phase transition with the low-temperature dual corresponding to a gas of particles in AdS and the high temperature corresponding to the AdS-Schwartzchild black hole (105) \cite{LectonGrav}. 

\begin{figure}[h]
		\begin{center}
			\includegraphics[keepaspectratio,scale=0.7]{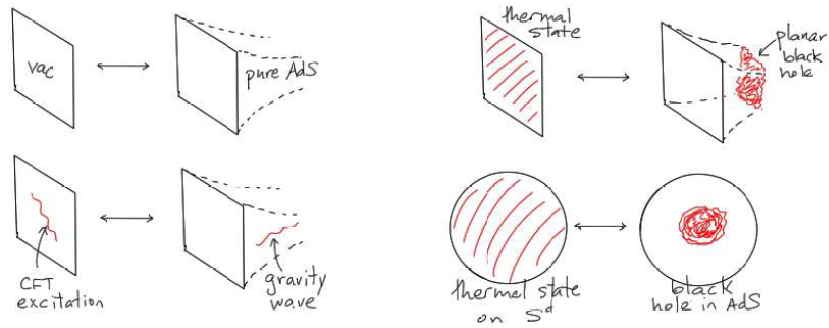}
		\end{center}
			\caption{Different CFT states correspond to different bulk geometries. Credits: \cite{LectonGrav}}
	\end{figure}

\newpage


	\section{Wormholes and Entanglement}

Bekenstein's work on black hole thermodynamics led to the definition of the concept of black hole entropy, given by

	\begin{equation}
		S=\frac{A}{4G}.
	\end{equation}
		
where $A$ is the black hole's surface area.
Nevertheless, the statistical interpretation of the black hole entropy is fully understood only in the context of the AdS/CFT correspondence as it allows for a complete definition of the underlying quantum state of the theory. The Schwartzchild-AdS is dual to a high temperature thermal state of the CFT on a sphere. This has a discrete spectrum of energies $\ket{E_i}$ and the thermal state corresponds to the canonical ensemble. The microstates counting for the entropy is, up to a numerical factor due to strong couplings in the CFT, matching with the black hole entropy and gives the correct relation between entropy and temperature. Therefore, the entropy of the thermal state of the CFT is dual to the area of the black hole horizon. This idea has been generalised in the Ryu and Takayanagi conjecture \cite{Ryu} which can be roughly stated as:

	\begin{description}
\item[\underline{Def}:] For any CFT state corresponding to some asymptotically AdS spacetime (with or without black hole), and for any subsystem of the CFT, the entropy of the subsystem corresponds to the area of a particular surface in the corresponding spacetime.
	\end{description} 

Since the entropy of quantum subsystems are best treated with the tools of quantum information theory, the conjecture draws the attention to the connection between quantum information and high energy/gravitational physics. Indeed, it is through the use of information theoretical techniques that the relation between entanglement and geometry will now be presented.

		\subsection{The Thermofield Double State}

Any ensemble of states  $\ket{\psi_i}$ of a quantum system can be represented with a density matrix 
	\begin{equation}
		\rho = \sum_i p_i\ket{\psi_i}\bra{\psi_i}.
	\end{equation}

Density matrices are an optimal tool to describe quantum systems which state is mixed or not completely specified. They can be though as operators on the Hilbert space of the theory obeying the properties:

	\begin{enumerate}
\item Non negativity:
		\begin{equation}
			\rho \geq 0.
		\end{equation}
\item Unit trace:
		\begin{equation}
			Tr[\rho]=1
		\end{equation}
\item Hermiticity:
		\begin{equation}
			\rho^\dag=\rho.
		\end{equation}
	\end{enumerate}

The expectation value of an operator $\hat{O}$ can be computed by taking its trace against the density matrix of the system
	\begin{equation}
		\langle\hat{O}\rangle = Tr[\rho \hat{O}].
	\end{equation}
Associated with each density matrix there is a Von Neumann entropy which quantifies the amount of information available in the system and its number of microstates
	\begin{equation}
		S(\rho) = -Tr[\rho log(\rho)]
	\end{equation}
For a \textit{canonical ensamble} or \textit{thermal state} representing a system weakly coupled with a thermal bath,  the density matrix can be obtained by maximising the entropy (119) subject to a fixed expectation value for the energy of the system 
	\begin{equation}
		Tr[\rho H]=E.
	\end{equation}
The resulting thermal density matrix takes the form
	\begin{equation}
		\rho = e^{-\beta H} = \frac{1}{Z} \sum_i e^{-\beta E_i}\ket{E_i}\bra{E_i}
	\end{equation}
where Z is the partition function 
	\begin{equation}
		Z= \sum_i e^{-\beta E_i}
	\end{equation}

and $\beta = T^{-1}$ is chosen to satisfy (121). 
If the $\ket{E_i}$ are identified with the CFT spectrum of energies and the temperature T is set to the Hawking temperature, then (122) represents the asymptotically AdS black hole dual to the  CFT. Indeed, a black hole carries the maximum amount of entropy for a given volume.
When dealing with quantum systems and subsystems, one has to consider that entanglement can enter the picture and needs to be accounted for. A common way of measuring the entanglement of a system is by measuring how much the system fails to factorise into a product state of subsystems

	\begin{equation}
		\ket{\psi} = \ket{\psi_A}\otimes \ket{\psi_B}.
	\end{equation}

When a state factorises as in (125), all the information about the subsystems is completely contained in each of them, and their entropy is minimal. Therefore, entanglement is equivalent to classical uncertainty about the state of the subsystems and can therefore be measured by the Von Neumann entropy, which in this case is called \textit{entanglement entropy}. This implies that a thermal canonical ensemble which maximises the entropy also maximises the entanglement between its subsystems.
When a system is in a mixed state $\rho$, it is possible to consider it as part of a bigger system in a pure state. This process is called \textit{purification} and it is achieved by doubling the original Hilbert space $\mathcal{H}_1$ such that $\mathcal{H}= \mathcal{H}_1 \otimes \mathcal{H}_2 $. The density matrix of a general purification can be written as

	\begin{equation}
		\ket{\Psi}= \sum_i \sqrt{p_i} \ket{\psi_i}_1\otimes \ket{\psi_i}_2,
	\end{equation}

where $\ket{\psi_i}_2$ is a set of orthogonal states in $\mathcal{H}_2$ which has a number of dimensions at least as big as the number of non-zero eigenvalues of $\rho$. This is named \textit{schimdt decomposition} and can be achieved starting from any $\rho$. For a canonical ensemble as in (123), the state can be purified using an exact copy of itself. The purification obtained in this way has the name of \textit{Thermofield Double State} and lives in the space $\mathcal{H}_A \otimes \mathcal{H}_B$

	\begin{equation}
		\ket{TFD} = \frac{1}{\sqrt{Z}}\sum_i e^{-\frac{\beta E_i}{2}} \ket{E_i}_A\ket{E_i}_B .
	\end{equation}
When one of the two system is traced out, the original thermal density matrix (122) is recovered. For example, by tracing out the $B$ part of the system
	\begin{equation}
		\rho_a= Tr_B[\rho_{tfd}] = \frac{1}{Z}\sum_i e^{-\beta E_i} \ket{E_i}\bra{E_i}_A \underbrace{Tr[\ket{E_i}\bra{E_i}_B]}_\text{1}.
	\end{equation}

Since the partial thermal density matrix maximises the entanglement entropy, this result shows that the two subsystems $A$ and $B$ are maximally entangled.
If one wants to purify the thermal state corresponding to a CFT, then it is necessary to take the CPT conjugate of the energy eigenbasis 

	\begin{equation}
		\ket{\tilde{E}_i} = \Theta \ket{E_i}.
	\end{equation}

In this way, the particles in the CFT 1, which will now be labelled \textit{left CFT}, will be entangled with their antiparticles in the CFT 2, which is now named \textit{right CFT}, and the thermofield double state can be written as

	\begin{equation}
		\ket{TFD} = \frac{1}{\sqrt{Z}}\sum_i e^{-\frac{\beta E_i}{2}} \ket{E_i}_L\ket{\tilde{E}_i}_R .
	\end{equation}

		\subsection{ER=EPR}

In \cite{Hartman2013} Maldacena and Hartman realised that the extended Schwartzchild-AdS black hole spacetime (figure 6) is best represented by the thermofield double state (129). Instead of representing two separate AdS spacetimes whose boundary CFTs are entangled in the TFD state, they represent two asymptotically AdS regions connected through the interior of their Schwartzchild black hole. Each copy of the CFT lives in one of the asymptotically conformal timelike boundaries at the left and right side of the diagram while the bulk contains an Einstein-Rosen bridge that connects the two black holes. This can be seen by studying correlation functions of primary operators belonging to different boundaries. Let $\mathcal{O}_L$ be an operator built from local fields on the left CFT

    \begin{equation}
        \mathcal{O}_L=\phi_L(x_L)\psi_L(y_L)...
    \end{equation}

and let $\mathcal{O}_R$ be an operator built from the right CFT local fields. Although the left and right systems are not coupled in the Lagrangian of the double system the 2 point correlator

    \begin{equation}
        \bra{TFD}\mathcal{O}_L\mathcal{O}_R\ket{TFD}
    \end{equation}

can be non zero. This happens because the two CFT are correlated through the entanglement of the thermofield double state. In the bulk, the correlators are non-zero because Witten diagrams can be drawn through the interior of the black hole. Since for high energies or massive fields, these correlators can be approximated by geodesics passing through the interior,  this would not be possible without the presence of a wormhole. Entangling the two CFT creates an Einstein-Rosen bridge that connects the two AdS spacetimes. The connection between entanglement and wormholes has been conjectured by Susskind to be a fundamental relation of reality and has been given the name ER=EPR \cite{ER=EPR}. The name of the conjecture refers to two papers, one by Einstein and Rosen describing wormholes for the first time, and the other one by Einstein Poldosky and Rosen about the nature of entanglement. The ER=EPR principle states that, whenever there is entanglement, there is also an Einstein Rosen bridge connecting the two systems. The bridge is usually too small to have a geometrical interpretation, but in the case of entangled black holes like in the AdS/CFT duality, it gains a classical geometrical structure and can be studied through the use of General Relativity.

		\subsection{The growth of the wormhole}

It may be useful to do a little recap of what has been said. By a careful choice of parameters, a thermal state of a conformal field theory defined on a sphere is dual to an AdS spacetime with a Schwartzchild black hole in equilibrium with its own Hawking radiation. If the thermal state is purified by doubling the CFT and putting the two theories in the entangled thermofield double state (129), the interiors of the corresponding black holes merge into an Einstein Rosen bridge. The width of the bridge is given by the minimal area surface separating the two spacetimes and corresponds to the area of the black hole's event horizon. Therefore, it is encoded in the entropy of the CFT. How does the state behave under time evolution?
In this setting, there are 2 different notions of time. The bulk time $\tau$, which is the coordinate time defining the Killing timelike symmetry of the static spacetime, runs forward in the right exterior and backwards in the left exterior. Inside the Einstein Rosen bridge, $\tau$ is spacelike and runs from left to right. Then, there are the boundaries time coordinates $t_l$ and $t_r$. They represent the time coordinate of the spacetime boundary in which the CFTs live and both run forward. 

	\begin{equation}
		\begin{array}{ccc}
		t_r & = & \tau \\
		t_l & = & -\tau
		\end{array}.
	\end{equation}

The generator of the Killing time translations is the Hamiltonian
	\begin{equation}
		H_{tot} = H_L - H_R.
	\end{equation}
The TFD state (129) is an eigenstate of this Hamiltonian and stays invariant under its action due to the phases cancelling out
	\begin{equation}
		\ket{TFD(\tau)} = e^{-i(H_Lt_l-H_Rt_r)}\ket{TFD}= \frac{1}{\sqrt{Z}}\sum_i e^{-\frac{\beta E_i}{2}} e^{-iE_i(t_l-t_r)}\ket{E_i}_L\ket{\tilde{E}_i}_R=\ket{TFD} .
	\end{equation}

\begin{figure}[h]
		\begin{center}
			\includegraphics[keepaspectratio,scale=0.55]{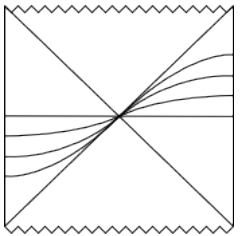}
		\end{center}
			\caption{The Hamiltonian $H_{tot}$ generates time translation that leave the TFD state invariant}
	\end{figure}

\newpage

 Each of the slices of figure 8 can be identified with the same geometry of the $\tau=t_l=t_r=0$ slice which cuts the Penrose diagram exactly in the middle.

\begin{figure}[h]
		\begin{center}
			\includegraphics[keepaspectratio,scale=0.6]{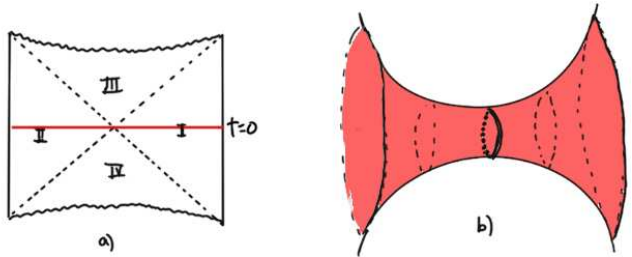}
		\end{center}
			\caption{a) Spacelike slice of the Penrose diagram indicating the TFD double state. b) Geometrical depiction of the Einstein-Rosen bridge connecting the AdS spacetimes. The circle in the middle shows the horizon as the minimal area surface connecting the two exteriors Credits: \cite{LectonGrav} }
	\end{figure}

It is by studying the behaviour of the wormhole under the translations generated by the Hamiltonian 
	\begin{equation}
		\tilde{H}_{tot} = H_L + H_R 
	\end{equation}
that an interesting behaviour emerges. $\tilde{H}_{tot}$ moves both $t_l$ and $t_r$ forward, therefore does not leave the TFD state invariant
	\begin{equation}
		\ket{TFD(t_l,t_r)} =  e^{-i\tilde{H}_{tot}t_{tot}}\ket{TFD}= \frac{1}{\sqrt{Z}}\sum_i e^{-\frac{\beta E_i}{2}} e^{-iE_i(t_l+t_r)}\ket{E_i}_L\ket{\tilde{E}_i}_R.
	\end{equation}

The interesting phenomenon occurs when the volume of the wormhole connecting the boundary states at $t_l$ and $t_r$ is measured as the Hamiltonian $\tilde{H}_{tot}$ acts. In order to do this, consider a spacelike slice that cuts the Penrose diagram from side to side and it is anchored at the boundary times $t_l$ and $t_r$. Every point in the curve is a (d-2) sphere and the curve itself is a codimension 1 hypersurface representing a spatial volume. This kind of surface fills the interior of the Einstein Rosen bridge. In order to pick the spacelike slice unambiguously, an assumption is made that the volume of these surfaces is always bounded from above. The spacelike slice is chosen to be the one that extremizes the volume, which is then guaranteed to exist and to be a maximum of the volume \cite{Stanford2014}.
In order to compute the volume of the slice, it is useful to understand the limit for which $t_l,t_r \rightarrow \infty$. The associated volume is represented by the light blue curve in figure 10. In this limit, the wormhole extends over an infinite range of $\tau$ and is hence invariant under $\tau$ translations. Each point of the Penrose diagram represents an $S^{d-2}$ and is therefore rotationally invariant. As a consequence, the maximum volume surface is located at a fixed value of r which volume per unit $\tau$ is

	\begin{equation}
		\frac{dV}{d\tau} = \Omega_{d-2}r^{d-2}\sqrt{|f(r)|}.
	\end{equation}

where 

	\begin{equation}
		f(r) = 1 - \frac{\mu}{r^{d-3}} + \frac{r^2}{l_{ads}^2}.
	\end{equation}

is taken from (105).

	\begin{figure}[h]
		\begin{center}
			\includegraphics[keepaspectratio,scale=0.7]{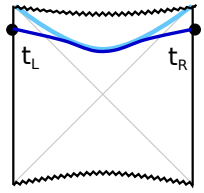}
		\end{center}
			\caption{The (d-1) volume surface (dark blue) probes the volume of the wormhole at boundary time $t_l$, $t_r$. When both the times are sent to infinity, the spacelike slice becomes extremal (light blue). Credits: \cite{Stanford2014}}
	\end{figure}

The RHS of expression (138) needs to be maximised over r in the inside of the wormhole, which spans from $r=0$ to $r=r_{Schwartz}$. The function (139) has a maximum at
	\begin{equation}
		r_m=\left( (d-3)\frac{\mu}{2} \right) ^\frac{1}{d-1}
	\end{equation}
the maximum volume is then denoted by $V_d$ 
	\begin{equation}
		V_d = \Omega_{d-2}r_m^{d-2}\sqrt{|f(r_m)|}.
	\end{equation}
The volume for the finite time Enstein Rosen bridge (dark blue line in figure 10) can be computed by considering that codimension 1 with (d-2) spherical symmetry at each point are geodesics of the metric 
	\begin{equation}
		ds^2 = -r^{2(d-2)}f(r)dt^2 + r^{2(d-2)}dr^2.
	\end{equation}
Parametrise the geodesic with a parameter $\lambda$ such that the volume is described by $r(\lambda)$ and $t(\lambda)$.
The Lagrangian approach with 
	\begin{equation}
		\mathcal{L} = -\frac{1}{2} r^{2(d-2)}f(r)(\dot{t})^2 +\frac{1}{2} r^{2(d-2)} (\dot{r})^2
	\end{equation}
where the dot represents a derivative with respect to $\lambda$, gives the conserved quantity associated with time translation 
	\begin{equation}
		E=r^{2(d-2)}\dot{t}.
	\end{equation}
The surface is a spacelike geodesic and must obey the parametrisation constraint
	\begin{equation}
		g_{\mu\nu}\frac{dx^\mu}{d\lambda}\frac{dx^\nu}{d\lambda}=1.
	\end{equation}
Using (144) one obtains the additional relation
	\begin{equation}
		\dot{r} = \frac{\sqrt{E^2 + r^{2(d-2)}f(r)}}{r^{2(d-2)}}.
	\end{equation}

Let the intersection points of the geodesic with the event horizon have Kruskal coordinates $(u_l,0)$ and $(0,v_r)$ so that the maximum volume surface connecting them measures the inside of the wormhole. Through the use of boost symmetry, set $u_l=v_r$

\begin{figure}[h]
		\begin{center}
			\includegraphics[keepaspectratio,scale=0.7]{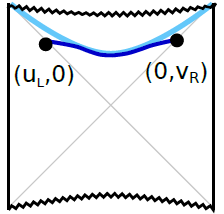}
		\end{center}
			\caption{The geodesics of (142) can be used to prove the volume of the wormhole at finite time. The intersection points are boosted so that $u_l=v_r$ Credits: \cite{Stanford2014}}
	\end{figure}
\newpage

The volume can now be computed and will be characterized by the energy E
	\begin{equation}
		V(E)= \int ds = \int \sqrt{g_{\mu\nu}\dot{x}^\mu\dot{x}^\nu}d\lambda= 2\int_{r_{turn(E)}}^{r_{Schw}} \frac{dr}{\dot{r}}=2\int_{r_{turn(E)}}^{r_{horizon}} \frac{r^{2(d-2)}}{\sqrt{E^2+r^{2(d-2)}f(r)}}dr
	\end{equation}

where $r_{turn}$ is the point at which the denominator vanishes.
The integral is regular near the upper limit. Instead of going through the calculation, it is insightful to notice that as E increases, the turning point approaches $r_m$ where the extremum of $r^{2(d-2)}f(r)$ lies, as shown in (140). A logarithmic divergence develops at this point as the surface starts running close to $r=r_m$ which is finite and positive. This implies that as the generator of the time translation (136) acts, the maximal volume surface through the wormhole hugs the limiting case given by the extremal surface at infinite time.
In the case of the unperturbed TFD state, the surface stays close to infite time limit until $\tau_l \approx - t_l$ and $\tau_r \approx t_r$. Therefore, the volume of the Einstein Rosen bridge for $|t_l+t_r| \gg \beta$ is given by

	\begin{equation}
		V(t_l,t_r) = V_d|t_l+t_r| + O(1)
	\end{equation}

where O(1) is an infinite but state- independent constant that regulates the divergence of the volume outside the wormhole.
Since the black hole dual is a high temperature CFT, $\beta$ will not be large.
Therefore, after a short transient, the volume of the bridge grows \textit{linearly} with the boundary CFT time.

\newpage
	\section{Complexity as Volume or Complexity as Action?}

		\subsection{Complexity of the Thermofield Double State}
The growth of the Einstein Rosen bridge classically goes on forever, at the same time the dual boundary theories reach thermal equilibrium very quickly after a scrambling time \cite{ComplexityBH}

	\begin{equation}
		t_*\sim\beta log(S)
	\end{equation}
 \newline

when all the evolution seems to stop. How can the growth of the wormhole then be described over a long period of time by the dual boundary theory? An equivalent question is asking if there are properties of the gauge theory wave function that can serve as clocks, and for how long will they be valid clocks? The truth is that the evolution does not actually stop, subtle quantum properties continue to evolve towards equilibrium after the system has been scrambled. There is a perfect candidate for a quantity that continues to grow linearly after the scrambling time and does so for a very long period. Quantum complexity.  The conjecture proposed is that the complexity of the TFD state which grows linearly with time under the time evolution generated by (136) is dual to the growth of the Einstein Rosen bridge connecting the two AdS black holes. By complexity of the TFD state, it is meant the complexity of the time evolution operator 

	\begin{equation}
		U(t_l,t_r,0,0)=e^{-i(H_Lt_l+H_Rt_r)}.
	\end{equation}

There are two different ways of relating complexity to the growth of the bridge, one with its volume, and the other with the action of a particular patch of spacetime.

		\subsection{Complexity as Volume}

The connection between the complexity of the boundary state and the volume of the wormhole was the first to be proposed \cite{Stanford2014} and can be stated as
	\begin{equation}
		\mathcal{C}(t_l,t_r) \approx \frac{V(t_l,t_r)}{l_{ads}G}
	\end{equation}
where $V(t_t,t_r)$ is defined as in (148). This equation has to be consider within a factor of proportionality of order 1 in order to account for the ambiguity in the definition of complexity.
To better see how this conjecture is supported, take a high temperature black hole $\mu \gg 1$ such that
	\begin{equation}
		r_m^{d-2}\sqrt{|f(r)|} = \frac{\mu}{2}.
	\end{equation}
Substituting into (141) gives
	\begin{equation}
		V_d= \frac{8\pi Gl_{ads}}{d-2}M.
	\end{equation}
Therefore, since $M\propto ST$
	\begin{equation}
		\mathcal{C}(t_l,t_r) \propto ST|t_l+t_r|.
	\end{equation}
For simplicity, consider an equal time translation on both CFT such that $t_l=t_r$ and $t_l+t_r=t_{tot}$, then
	\begin{equation}
		\frac{d\mathcal{C}}{dt_{tot}} \propto ST
	\end{equation}

In order to connect this result with the rate of growth of quantum complexity defined in (20) and (83), it is necessary to define a relation between the circuit time and the CFT/black hole time. The circuit time of (20) and (83) is a dimensionless parameter. For all non-extremal black holes, there exists a dimensionless time parameter that describes the near-horizon geometry. Consider the Rindler coordinates

	\begin{equation}
		ds^2= -\rho^2d\tau^2+d\rho^2+r(\rho)^2d\Omega^2
	\end{equation}

where $\rho$ is the proper distance from the horizon and $\tau$ is the dimensionless hyperbolic boost angle. One can recall that the circuit model for the black hole mentioned in chapter III consists of non-localised qubits living in the stretched horizon which dynamics are regulated by the 2-local Hamiltonians of the type (15). Therefore, the natural candidate for the circuit time is $\tau$. The computation is red-shifted at infinity where the boundary time regulates the evolution of the CFT. For non-extremal black holes, the asymptotic time and the Rindler time are related by

	\begin{equation}
		\tau = \frac{2\pi t}{\beta}=2\pi Tt.
	\end{equation}

Hence, by remembering that to model the internal dynamics of a black hole the number of qubits necessary is of the order of the entropy (77), the rate of growth of circuit complexity at the black hole horizon and quantum complexity at the CFT boundary match

    \begin{equation}
        \frac{d\mathcal{C}}{dt} =ST \propto \frac{K}{2} = \frac{d\mathcal{C}}{d\tau}.
    \end{equation}

The complexity-volume duality (C=V)  has a number of nice features. The maximal spacelike slice is a robust geometric object and its growth is naturally proportional to $TS$. The duality works as $TS$ is a rough estimate of the rate of complexification of the boundary state. If the theory is probed by the insertion of precursors (21), the gravity dual is represented by a shockwave geometry and calculations using the C=V conjecture have matching cancellations representing the switchback effect appropriately \cite{Stanford2014}.

		\subsection{Lloyd's bound on complexity growth}

Before talking about the other conjecture relating the growth of the wormhole with quantum complexity, it is necessary to better analyse the rate of complexity growth. In the same way as Bekenstein's entropy is a limit on the maximum amount of information that can be stored in a given volume of space, an upper bound exists on how fast the complexity of a quantum system can grow. Previous bounds, like the Aharonov-Anandan-Bohm, bound, where found for the time that a system takes to reach an orthogonal state. These bounds are dependent either on the standard deviation of the energy or on its average above the ground state. Lloyd's noted that N parallel copies of a computer could be understood to compute N times as fast and that the expected energy grows as N while the standard deviation only grows as $\sqrt{N}$. Henceforth, he conjectured that the rate of computation is proportional to the average energy. When this idea is applied to complexity growth, the number of simple gates needed to prepare the state from a reference state of an isolated unitarily evolving quantum system is

    \begin{equation}
        \frac{d(gates)}{dt} \leq \frac{2E}{\pi\hbar}
    \end{equation}

When talking about continuum field theories, a more refined notion of complexity is necessary. This notion should preserve all the characteristic features of complexity such as the linear growth under local Hamiltonian evolution and its proportionality to the number of active degrees of freedom. When talking about complexity growth, of the two sided Penrose diagram, a good reference state is the thermofield double state. If $E_\psi$ is taken to be the average energy of the state $\ket{\psi}$ with respect to the ground state, then Lloyd's bound for suitable semiclassical bulk states and dual field theories becomes

	\begin{equation}
		\frac{d\mathcal{C}}{dt}[e^{-iHt}\ket{\psi}] \leq \frac{2E_\psi}{\pi\hbar}.
	\end{equation}
When the bulk state has a black hole, the bound takes the form that will be relevant for discussing the complexity-action conjecture
	\begin{equation}
		\frac{d\mathcal{C}}{dt} \leq \frac{2M}{\pi\hbar}.
	\end{equation}

		\subsection{Complexity as Action}

The C=V conjecture has some imperfections. For small black holes, $l_{ads}$ in (151) is not an appropiate lenght scale and needs to be substituted with $r_{Schwartz}$. Moreover, the matching with the CFT complexity is only approximate and the maximal spacelike slices do not foliate the entire geometry behind the black hole horizon as their infinite time limit is the slice characterized by (141).
More recently, another conjecture has been proposed in \cite{CAction} that relates the complexity of the boundary state with the action of a particular patch in the bulk spacetime.
The idea for the C=A conjecture was born from the realisation that when the C=V relation is multiplied and divided by the $l_{ads}$ scale, the numerator becomes the \textit{world volume} of the Einstein Rosen bridge

    \begin{equation}
        \mathcal{C} \sim \frac{V}{l_{ads}G} \sim \frac{\mathcal{W}}{l_{ads}^2G}
    \end{equation}

where $\mathcal{W} \equiv Vl_{ads}$ has units of spacetime volume. From (97) it is possible to see that $\frac{-1}{l_{ads}^2}$ is proportional to the cosmological constant $\Lambda$, which is roughly the classical action of the world volume $\mathcal{W}$. To precisely define the complexity-action conjecture, define first the Wheeler-DeWitt (WDW) patch to be the spacetime region determined by the data on any spacelike slice within $\mathcal{W}$ which is naturally associated with the boundary state. The WDW patch given the boundary CFT times is $\mathcal{W}(t_l,t_r)$ and can be thought as the union of all the spacelike surfaces anchored at $t_l$ and $t_r$. It is the spacetime region trapped between the forward and backward light rays sent from the boundaries at $t_l$ and $t_r$.

\begin{figure}[h]
		\begin{center}
			\includegraphics[keepaspectratio,scale=0.7]{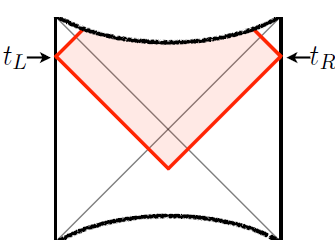}
		\end{center}
			\caption{Wheeler-DeWitt patch $\mathcal{W}(t_l,t_r)$ Credits: \cite{CAction}}
	\end{figure}
\newpage

It needs to be pointed out that $\mathcal{W}$ is not a causal patch and cannot be all monitored by a single observer which is consistent with complexity not being an observable in quantum mechanics. If suitable boundary terms on $\partial\mathcal{W}$ are imposed, the complexity $\mathcal{C}$ can be thought as the minimum number of gates from the chosen universal set necessary to produce the state $\ket{TFD(t_l,t_r)}$ and the complexity-action duality is defined as

    \begin{equation}
        \mathcal{C} [\ket{TFD(t_l,t_r)}]= \frac{\mathcal{A}_{\mathcal{W}}}{\pi\hbar}.
    \end{equation}

The action of the WDW patch for an Einstein-Maxwell theory is

    \begin{equation}
        \mathcal{A}_\mathcal{W} = \frac{1}{16\pi G}\int_\mathcal{W} \sqrt{|g|}(\mathcal{R} - 2\Lambda) - \frac{1}{16\pi}\int_\mathcal{W}\sqrt{|g|}F_{\mu\nu}F^{\mu\nu}+ \frac{1}{8\pi G}\int_{\partial\mathcal{W}} \sqrt{|h|}K,
    \end{equation}

which contains the Einstein-Hilbert action with negative cosmological constant, the Maxwell term and the boundary York, Gibbons, Hawking surface action where $K$ is the extrinsic curvature tensor trace. The convention for the extrinsic curvature used is that spacelike normals point outwards and timelike normal inwards.
As an example, the action of the WDW patch will be computed for a static neutral black hole following the approach of \cite{Brown2016}, where computations for general Reissner-Nordstrom-AdS black holes and rotating BTZ black holes can be found.
The metric for the AdS-Schwartzchild is 

    \begin{equation}
        ds^2 = -f(r)dt^2 + \frac{dr^2}{f(r)} + r^2d\Omega^2_{d-2}
    \end{equation}

with $f(r)$ as in (139). The volume of the WDW patch is infinite but time-independent due to the translation symmetry outside the black hole. As time passes $t_l$ and/or $t_r$ move forward, the patch grows in some places and shrink in others. Some of the spacetime region trapped inside the WDW patch belongs to the past horizon, but at late times, when $|t_l+t_r|\gg\beta$, it shrinks exponentially to zero. The patch has 2 dimensions and a $S^{d-2}$ sphere at each point which has constant size up to terms that are exponentially small in time. The Gauss-Bonnet theorem can be used on the 2 dimensions that define a manifold without boundary. The theorem states that 

    \begin{equation}
        \int_M k dA = 2\pi \chi(M)
    \end{equation}

where $k$ is the Gaussian curvature of the surface and $\chi(M) = 2-2g$ is its Euler's characteristic, $g$ being the genus. The contribution of this bit is therefore topological and must be independent of time.
Therefore, at late times, the only relevant contribution comes from the region of the Wheeler-DeWitt patch that is contained inside the future horizon.

\begin{figure}[h]
		\begin{center}
			\includegraphics[keepaspectratio]{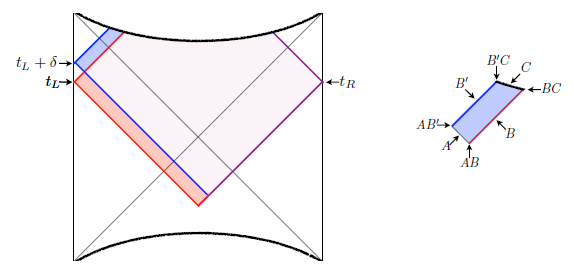}
		\end{center}
			\caption{Uncharged AdS black hole WDW patch as the boundary time $t_l$ increases infinitesimally. Credits: \cite{Brown2016}}
	\end{figure}
\newpage

As it can be seen from (figure 13), the only changes in the action come from the A ($r=r_{Schwartz}$) and C ($r=0$) segments. This happens because the contribution from $B$ is replaced by that at $B^\prime$, but since they are the same under time translation symmetry, the action is not affected. The contribution lost from the corner $AB$ is replaced by $AB^\prime$ and the same happens for $BC$ and $B^\prime C$.
In order to compute the bulk contribution, recall (97) 
    \begin{equation}
        \Lambda = \frac{-(d-1)(d-2)}{2l_{ads}^2}.
    \end{equation}
and contract (94) to get
    \begin{equation}
        \mathcal{R} = \frac{2d}{(d-2)}\Lambda
    \end{equation}
the volume element of (162) is
    \begin{equation}
        \sqrt{|g|}dx^d= r^{d-2}drd\Omega_{d-2}
    \end{equation}
where $d\Omega_{d-2}$ is the volume element of a {d-2} unit sphere.
Hence, the Einstein-Hilbert term of (162) becomes
    \begin{equation}
        \frac{d\mathcal{A}_{EH}}{dt_l} = \frac{1}{16\pi G}\int_\mathcal{W} \left( \frac{2d}{(d-2)}-2 \right)  \underbrace{ \left( -\frac{(d-1)(d-2)}{2l_{ads}^2} \right) }_\text{$\Lambda$} r^{d-2}drd\Omega_{d-2}= -\frac{r_{schw}^{d-1}\Omega_{d-2}}{8\pi G l_{ads}^2}.
    \end{equation}
The integral goes all the way to the singularity, but the contribution received near r=0 is minimal.
The boundary integral is carried out by expressing the trace of the extrinsic curvature of a constant r-surface as 
    \begin{equation}
        K= \frac{1}{2} n^r \frac{\partial_r(r^{2(d-2)}f(r))}{r^{2(d-2)}f(r)}
    \end{equation}
giving a contribution for the YGH term of
    \begin{equation}
        \frac{d\mathcal{A}_{ygh}}{dt_l} = \left[ -\left(\frac{d-1}{d-2}\right)M + \frac{r^{d-3}\Omega_{d-2}}{8\pi G}\left( (d-2) + (d-1)\frac{r^2}{l_{ads}^2} \right) \right]^{r_{schwartz}}_{0}
    \end{equation}
Combining (167) and (169) with the fact that $f(r_{schwartz})=0$
    \begin{equation}
        \frac{d\mathcal{A}_{\mathcal{W}}}{dt_l}=2M
    \end{equation}
which is a powerful general result for neutral static AdS black holes of any size.
According to the conjecture, Complexity=Action which translates into a rate of change of complexity of
    \begin{equation}
        \frac{d\mathcal{C}}{dt}=\frac{2M}{\pi\hbar}.
    \end{equation}
Therefore, AdS-Schwartzchild black holes saturate Lloyd's bound on complexity growth (161) whatever their size and number of spacetime dimensions. The implications are not only that the Complexity-Action conjecture is valid, at least in this case, but also that neutral static black holes are the fastest computers in the universe. This is a much general and precise derivation with respect to the Complexity=Volume duality and hence a much more powerful result.

		\subsection{Exponential time breakdown of General Relativity}

Regardless of which conjecture is believed to be the correct one, the connection between complexity and General Relativity highlights a problem in the classical description of the wormhole interior. According to GR, the Einstein Rosen bridge grows indefinitely. If the growth is dual to the complexification of the boundary state, it should stop once the complexity has reached its maximum. Moreover, a problem occurs when the recurrence time for complexity is reached. The complexity should undergo a violent fluctuation that should reduce it to a value close to the minimum. There is no classical equivalent for this behaviour since the wormhole growth is classically stable and linear. At double exponential time in the degrees of freedom of the black hole, a breakdown of General Relativity occurs and a complete quantum gravity theory may be necessary to explain the physical behaviour of the system.

\newpage
	\section{Summary and Discussion}

The aim of this work was to delineate the relationship between a triangle of systems through which General  Relativity, Quantum Field Theory and Quantum Information Theory are connected.

	\begin{enumerate}
		\item 		$K$ qubits interacting through a k-local Hamiltonian ($\mathcal{Q}$)
			
		\item 		Space of unitary operators acting on $K$ qubits ($\mathcal{A}$)
			
		\item		Thermofield double state of two entangled CFT with AdS-Schwartzchild dual spacetimes connected through an Einstein Rosen bridge
	\end{enumerate}

The connection between $\mathcal{Q}$ and $\mathcal{A}$ happens through the notion of quantum complexity. The idea of quantum complexity arises in the purely quantum system $\mathcal{Q}$ as a measure of the operational distance between two quantum states or unitary operators. The $K$ qubits start in a reference state with no correlations and with a complexity equal to zero. When defined correctly, relative complexity becomes a perfect candidate for a metric on a Riemannian manifold. The role of the manifold is played by $\mathcal{A}$, the Lie group of unitary operators $SU(2^K)$. The complexity metric is then defined as in (49) and (50). After this imposition, $\mathcal{A}$ becomes negatively curved and the distance between any two points, which represent unitary operators, becomes their relative complexity. The time evolution of a unitary operator through $\mathcal{A}$  can be represented as the motion of a free non-relativistic particle. If the Hamiltonian chosen to dictate this motion is the same that regulates the dynamics of $\mathcal{Q}$, the evolution of a free particle starting from the identity can be made to represent the complexification of the time evolution unitary operator that evolves the state of the $K$ qubits. This relation allows for the identification of the positional entropy of $\mathcal{A}$ with the complexity of $\mathcal{Q}$. Therefore, a II law of quantum complexity is defined. Complexity grows with time as a statistical consequence of the fact that the number of accessible unitary operators grows exponentially with the complexity. It grows linearly and saturates after a time that is exponential in the number of qubits. The Poincar\'e recurrence occurs, instead, after a time that is doubly exponential in the size of $\mathcal{Q}$ (figure 5).
\newline
\newline
At the same time, $\mathcal{Q}$ and the thermofield double state of the entangled AdS/CFT are also connected by the concept of complexity. The properties that complexity exhibits, like the scrambling time and the switchback effect, are due to the k-local Hamiltonian used to determine the dynamics of the $K$ qubits. This Hamiltonian comes from modelling the degrees of freedom of a black hole as qubits living on a stretched horizon, their number is proportional to the entropy and hence the area of the black hole. The AdS/CFT duality allows for the identification of the CFT entropy with the entropy of the black hole. The duality is supposed to map all the bulk quantities and behaviours to the boundary CFT. Therefore, when General Relativity predicts that the wormhole connecting the black holes interior grows linearly with time, the growth of the quantum complexity of the thermofield double state is used as a boundary dual. This leads to two conjectures. The first relates the complexity of the TFD state with the volume of the wormhole and the second with the action of a particular spacetime patch called Wheeler-DeWitt patch. The complexity-action conjecture allows for the realisation that neutral static black holes are not only the most efficient memories in nature but also the fastest computers.
\newline
\newline
The only link missing is the relation between the classical system $\mathcal{A}$ and the AdS/CFT duality. This relation has not yet been defined but a couple of things can be said. If the second law of quantum complexity holds also for the CFT, then the growth of the wormhole can also be explained as a statistically overwhelming phenomenon. This would delineate another connection between General Relativity and statistical mechanics. The growth of the Einstein Rosen bridge is nothing less than the creation of spacetime. This sparks the idea of the emergence of spacetime as a statistical phenomenon prompted by subtle correlations between quantum fields. The connection between spacetime and entanglement has already been established \cite{VANRAAMSDONK2010}. Entanglement between quantum fields holds spacetime together. Therefore, the development of quantum correlations like complexity may be another factor contributing to its emergence.
\newline
\newline
Whatever the precise interplay between all these quantities is, something is sure. Since the holographic principle has been formulated, there has been an ever-growing connection between quantum information, entanglement and the structure of spacetime. Many details about this relation have still to be understood and refined, but the fact that the fundamental constituents of reality can be understood in terms of qubits and quantum computation techniques is, to say the least, astounding. It will be fascinating to see how all of this will come into play in the realisation of the long searched theory of quantum gravity.


	\section{Acknowledgements}

I am really thankful to Dr Johannes Bausch for accepting the supervision of this work and for all the help he gave me throughout the process of writing this thesis.
\newline
\newline
Many thanks to my dear friends and colleagues Enrico Fontana and Nicol\'o Primi. Thanks for listening to my complaints about the amount of work that went into this dissertation and for cheering me up all the time. 

	\newpage

\bibliographystyle{plain}
\bibliography{Essayreference}

\end{document}